\documentclass{emulateapj}
\usepackage{graphicx}
\usepackage{natbib}

\def\apjref#1;#2;#3;#4 {\par\pni\ #1,  #2, {\bf #3}, #4. \par}

\def\simlt{\lower.5ex\hbox{$\; \buildrel < \over \sim \;$}}
\def\simgt{\lower.5ex\hbox{$\; \buildrel > \over \sim \;$}}
\def\solar{\ifmmode _{\mathord\ odot}\else $_{\mathord\odot}$\fi}
\def\msun{\ifmmode {\rm M}_{\mathord\odot}\else $M_{\mathord\odot}$\fi}
\def\lsun{\ifmmode {\rm L}_{\mathord\odot}\else $L_{\mathord\odot}$\fi}
\def\xpvec{x\lla*p{\lower1ex\hbox{$\scriptstyle\sim$}}_{\scriptscriptstyle\perp}}
\def\xvec{x\llap{\lower1ex\hbox{$\scriptstyle\sim$}}}
\def\qvec{q\llap{\lower1ex\hbox{$\scriptstyle\sim$}}}
\def\to{\ifmmode \rightarrow\else $\rightarrow$\fi}
\def\Mc{M\raise0.5ex\hbox{c}}
\def\del{\nabla}

\def\none{\ifmmode ^{-1}\else $^{-1}$\fi}
\def\two{\ifmmode ^{2}\else $^{2}$\fi}
\def\ntwo{\ifmmode ^{-2}\else $^{-2}$\fi}
\def\three{\ifmmode ^{3}\else $^{3}$\fi}
\def\nthree{\ifmmode ^{-3}\else $^{-3}$\fi}
\def\four{\ifmmode ^{4}\else $^{4}$\fi}
\def\nfour{\ifmmode ^{-4}\else $^{-4}$\fi}
\def\five{\ifmmode ^{5}\else $^{5}$\fi}
\def\nfive{\ifmmode ^{-5}\else $^{-5}$\fi}

\def\g{\ifmmode {\rm g}\else g\fi}
\def\kg{\ifmmode {\rm kg}\else kg\fi}

\def\cm{\ifmmode {\rm cm}\else cm\fi}
\def\m{\ifmmode {\rm m}\else m\fi}
\def\km{\ifmmode {\rm km}\else km\fi}
\def\pc{\ifmmode {\rm pc}\else pc\fi}
\def\ly{\ifmmode {\rm ly}\else ly\fi}
\def\au{\ifmmode {\rm au}\else au\fi}

\def\s{\ifmmode {\rm s}\else s\fi}
\def\Hz{\ifmmode {\rm Hz}\else Hz\fi}
\def\y{\ifmmode {\rm y}\else y\fi}

\def\K{\ifmmode {\rm K}\else K\fi}
\def\ster{\ifmmode {\rm ster}\else ster\fi}
\def\erg{\ifmmode {\rm erg}\else erg\fi}
\def\dyn{\ifmmode {\rm dyn}\else dyn\fi}

\renewcommand{\baselinestretch}{1.0}\small\normalsize

\begin{document}

\title{The Effects of Radiative Transfer on Low-Mass Star Formation}

\author{Stella S. R. Offner} \affil{Department of Physics, University of
California, Berkeley, CA 94720} 
\email{soffner@berkeley.edu}

\author{ Richard I. Klein}
\affil{Department of Astronomy, University of California, Berkeley 
CA 94720, USA, and Lawrence Livermore National Laboratory, P.0. Box 808, L-23,
Livermore, CA 94550, USA}

\author{Christopher F. McKee}
\affil{Departments of Physics and Astronomy, University of California, Berkeley, CA 94720, USA}

\author{Mark R. Krumholz }
\affil{ Department of Astronomy, University of California, Santa Cruz, CA 95064} 

\begin{abstract}
Forming stars emit a substantial amount of radiation into their natal environment.
We use ORION, an adaptive mesh refinement (AMR) three-dimensional gravito-radiation-hydrodyanics code, to simulate low-mass star formation in a turbulent molecular cloud.
We compare the distribution of stellar masses, accretion rates, and temperatures in the cases with and without radiative transfer, and we demonstrate that radiative feedback has a profound effect on accretion, multiplicity, and mass by reducing the number of stars formed and the total rate at which gas turns into stars. 
We also show, that once star formation reaches a steady state, protostellar radiation is by far the dominant source of energy in the simulation, exceeding viscous dissipation and compressional heating by at least an order of magnitude.
Calculations that omit radiative feedback from
protstars significantly underestimate the gas temperature and the strength
of this effect.
Although heating from protostars is mainly confined to the protostellar cores, we find that it is sufficient to suppress disk fragmentation that would otherwise result in very low-mass companions or brown dwarfs. 
We demonstrate that the mean protostellar accretion rate increases with the
final stellar mass so that the star formation time is only a weak function of mass.

\end{abstract}

\section {INTRODUCTION}

On large scales molecular clouds are generally observed to have limited temperature variations, a characteristic that results from the high efficiency of radiative cooling at typical cloud densities.  Consequently, simulations of molecular clouds frequently assume constant gas temperature, a convenient approximation for investigations of gas-dynamics, turbulence, and gravitational collapse \citep{gammie03, bonnell03, li04, tilley04, vazquez08}.
However, an isothermal assumption necessarily neglects the influence of heating due to gas compression, accretion, and stellar sources. 

The importance of the local gas temperature to the star formation process is motivated analytically when considered in combination with gravity.
The characteristic fragmentation scale for self-graviting gas of density, $\rho$, and sound speed, $c_{\rm s}$, is  given by the Jeans length:
\begin{equation}
\lambda_J =  \sqrt{{{\pi c_{\rm s}^2 } \over {G \rho}}} \propto \left({{T}\over{\rho}}\right)^{1/2}.
\end{equation}
Thus, for lower temperatures, gas is prone to gravitational instability at lower densities. 
In rotating self-gravitating disks, the criterion may be phrased in terms of the local column density $\Sigma$:
\begin{equation}
Q =  {{\kappa_{\epsilon} c_{\rm s}} \over {\sqrt{\pi G \Sigma}}} \propto T^{1/2},
\end{equation}
where the onset of gravitational instability occurs as the Toomre parameter, $Q$, approaches one and $\kappa_{\epsilon}$ is the epicyclic frequency.
Cold protostellar disks more readily develop spiral structure and become Toomre unstable, influencing protostellar accretion and driving fragmentation \citep{kratter08}. Overproducing low-mass objects or brown dwarfs in the stellar initial mass function is one side-effect of increased fragmentation \citep{bate09b, krumholzkm07}. 

Gas eventually becomes optically thick at densities orders of magnitude above the molecular cloud mean, and radiative cooling is no longer efficient. 
To investigate this transition, \citet{masunaga98} modeled a spherically symmetric core collapse including angle-dependent multi-frequency radiative transfer, resolving scales down to the accretion shock. They halted the calculation at the end of the first collapse phase, prior to the dissociation of $H_2$ and before protostellar feedback commences. The authors reported a characteristic transition density of $\sim 10^{-13}$ g cm$^{-3}$ for initially 10 K gas, above which the temperature increased with increasing density. In many turbulent simulations rudimentary heating due to gas compression is frequently represented using an equation of state \citep{li03, bate03, bate05, jappsen05, bonnell06, Offner08a, clark08, bate09a}. Although such an equation typically fits a more exact radiative transfer solution like that reported by \citet{masunaga98}, it neglects the instantaneous mean free path, multi-dimensional effects, dust chemistry, and time dependence of stellar sources.
 In fact, the subsequent paper, \citet{masunaga00}, demonstrated that gas temperatures may become significantly warmer as a result of protostellar feedback and that the temperature distribution is quite sensitive to the accretion luminosity.

To compromise between physics and computational expense, a few hybrid methods include heating
by solving explicit diffusion approximations, estimating the instantaneous radiative cooling, or extrapolating from previously tabulated temperatures \citep{stamatellos07, banerjee07, bonnell08}. 
Such methods are computationally cheaper and reproduce radiative heating for simple geometries. However, the suitability of these approximations is unclear for radiative problems involving clustered star formation in a turbulent medium, where the problem is highly non-linear, involves complex geometry, and the column density may not be a good indicator of the cooling rate. 
In addition, many of these approaches also neglect heating by stellar sources, which are crucial as we will show in this paper.
The unknown accuracy of radiation approximations and the deficiencies in handling temporal and spatial variations motivates our use of a full radiative transfer method, albeit one based upon the gray flux-limited diffusion approximation.

Relatively few authors have pursued 3D calculations including radiative transfer. 
These authors always adopt the flux-limited diffusion approximation and assume that the radiation field is frequency-independent, i.e., gray.
By modeling star formation with gray flux-limited diffusion (GFLD), it has been shown that a barotropic or polytropic equation of state (EOS) can underestimate the true heating at high densities even for simple, non-turbulent collapse problems \citep{boss00, whitehouse06}.  
The issue of radiative feedback is particularly acute for high-mass stars, which emit prodigious luminosities while forming \citep{krumholz06, krumholzkm07, krumholz09}.  To explore this point, \citet{krumholzkm07} contrasted simulations of collapsing, turbulent high-mass cores using an isothermal EOS to those using GFLD radiation transfer. 
The authors demonstrated that simulations with radiative transfer are able to produce a massive star formed from gas accretion, while barotropic or isothermal calculations may only produce a massive star via mergers of many smaller bodies.
Comparisons of the temperature distribution, assuming a barotropic EOS in lieu of radiative transfer, showed significant underestimation of the volume of heated gas and a much lower local maximum gas temperature.   


In the regime of low-mass star formation, \citet{bate09b} modeled several small clusters forming low-mass stars with the SPH radiative transfer method developed by \citet{whitehouse06}. The author compared these with previous published simulations using identical initial conditions and a barotropic EOS. The calculations including radiation transfer showed a substantial decrease in the number of brown dwarfs from 50\% of the number of objects to $<10$\%. 
This agrees with the prediction of \citet{matznerlevin05}, who assert incorrect disk fragmentation may produce  brown dwarfs  if irradiation is not included.
Accretion luminosity, which is emitted at the protostellar surface, generates a significant portion of the luminosity during the early stages of protostar formation.  Indeed, \citet{bate09b} found increased heating and fewer brown dwarfs at higher resolution but reported little difference in the final stellar distribution for resolutions of 0.5 AU versus 5.0 AU. Since 0.5 AU is much larger than protostellar radii, significant accretion heating was neglected. As \citet{bate09b} also neglected deuterium burning, the calculations represent a lower limit on the effects of radiative feedback.

In this paper, we model the formation of low-mass stars in a turbulent molecular cloud including GFLD using the ORION adaptive mesh refinement (AMR) code.  
We address the issue of radiative feedback, including all the important energy sources, and how it influences low-mass star formation.
Our study differs from previous work
in that we use source terms to account for accretion luminosity down to the stellar surface and include a stellar evolutionary model \citep{tan04}. 
We contrast a GFLD simulation to one without radiative transfer. We also perform a less time evolved calculation with resolution eight times higher to characterize the dependence of the solutions with and without radiative transfer on resolution.  We describe our method in \S 2. In \S 3, we compare and contrast the four simulations. 
We discuss caveats to our method in \S 4 and summarize our conclusions in \S 5. 
Comparisons to
observations will appear in a subsequent paper.

\section{METHODOLOGY AND INITIAL CONDITIONS \label{meth}}

\subsection{Numerical Methods}

For the purpose of comparison, we perform two calculations with identical resolutions and characteristic parameters. The first, which we denote RT, includes radiative transfer and feedback from stellar sources. The second, henceforth NRT, uses an EOS to describe the thermal evolution of the gas.
We perform both simulations using the parallel AMR code, ORION.
ORION utilizes a conservative second order Godunov scheme to solve the equations of 
compressible gas dynamics \citep{truelove98, klein99}: 
\begin{eqnarray}
{{\partial \rho} \over{ \partial t}} + \nabla \cdot (\rho {\bf v}) &=& 0, \\
{{\partial (\rho {\bf v})} \over{ \partial t}} + \nabla \cdot (\rho {\bf vv}) &=& -\nabla P - \rho \nabla \phi, \\
{{\partial (\rho e)} \over{ \partial t}} + \nabla \cdot [(\rho e + P){\bf v}] &=& \rho {\bf v} \nabla \phi - \kappa_R \rho (4\pi B -cE), 
\end{eqnarray}
where $\rho$, $P$, and $\bf v$ are the fluid density, pressure, and velocity, respectively. The total fluid energy is given by $e= 1/2 \rho {\bf{v}}^2 + P/(\gamma-1)$, where $\gamma=5/3$ for a monatomic ideal gas\footnotemark. The total radiation energy density is denoted by $E$, and $B$ is the Planck emission function.
ORION solves the Poisson equation for the gravitational potential $\phi$ using 
multi-level elliptic solvers with multi-grid iteration: 
\begin{equation}
{\nabla}^2 \phi = 4 \pi G  [ \rho + \sum_n m_n \delta({\bf x}-{\bf x}_n) ],
\end{equation}
where $m_n$ and ${\bf x}_n$ are the mass and position of the nth star.

\footnotetext{Most of the volume of the domain is too cold to excite any of the $H_2$ rotational or vibrational degrees of freedom, and thus the gas acts as if it were monatomic.}

ORION solves the non-equilibrium flux-limited diffusion equation using a parabolic solver with multi-grid iteration to determine the radiation energy density  \citep{krumholzkmb07}:
\begin{equation}
{{\partial E} \over{ \partial t}} - \nabla \cdot (\frac{c \lambda }{ \kappa_{\rm R} \rho} \nabla E) = \kappa_{\rm P} \rho (4 \pi B - cE) + \sum_n L_n W({\bf x}-{\bf x}_n) \label{radenergy}, 
\end{equation}
where $\kappa_{\rm R}$ and $\kappa_{\rm P}$ are the Rosseland and Planck dust opacities, and $L_n$ is the luminosity of the nth star. $W$ is a weighting function that determines the addition of the stellar luminosity to the grid (see Appendix A for details of the star particle algorithm). The flux-limiter is given by $\lambda = {1\over R} (\coth R -  {1\over R})$, where $R = |\nabla E/ (\kappa_{\rm R} \rho E) |$ \citep{levermore81}. 
%

We assume that the dust grains and gas are thermally well-coupled, an approximation we discuss further in Section 4.2. We obtain the dust opacities from a linear fit of the \citet{pollack94} dust model, which includes grains composed of silicates, trolites,  metallic iron, organics, and H$_2$O ices.
For gas temperatures $10 \leq T_g \leq 350$ K, the linear fit is given by:
 \begin{eqnarray}
 \kappa_R &=& 0.1 + 4.4 (T_g/350 ) \mbox{  cm$^2$ g$^{-1}$}, \\
 \kappa_P &=& 0.3 + 7.0 (T_g/375 ) \mbox{  cm$^2$ g$^{-1}$}.
 \end{eqnarray}
 These fits give $\kappa_{\rm R} =  0.23$ cm$^2$ g$^{-1}$ and $\kappa_{\rm P} =  0.49$ cm$^2$ g$^{-1}$ at the minimum simulation temperature, 10 K.
 Work by \citet{semenov03} explores the effect of dust composition, porosity, and iron content on the Planck and Rosseland average opacities. For the different models, they find a spread of more than an order of magnitude in the opacity at 10 K. The simplest model, based upon the assumption that dust grains are homogenous spheres, produces the lowest value for the Rosseland opacity, $\kappa_R \simeq 0.02$ cm$^2$ g$^{-1}$, while the most porous and non-homogenous grain models produce 10 K opacities as large as $\kappa_R \simeq  0.7$ cm$^2$ g$^{-1}$.  For temperatures above 100 K, the different dust models are more converged and the opacities are generally within a factor of 2. As a result, the temperature range from 10 K to 100 K is the most sensitive to dust assumptions. In this range, homogenous models increase roughly quadratically with temperature, while fluffier grain models increase linearly.
Our opacity fits are then close to the mean value of $\kappa_R =  0.16$ for the \citet{semenov03} models, although this value is more representative of porous and aggregate grains. As a result, our dust model is reasonable for the higher density regions of $n \gtrsim 10^7$ cm$^3$ typical of protostellar cores, but we may overestimate the dust opacity in the lower density cold gas by as much as a factor of 10. 

In studies of low-mass star formation, it is reasonable to neglect pressure exerted by the radiation field on the dust and gas. This is because the 
advection of radiation enthalpy is small compared to the rate the radiation diffuses through the gas:
\begin{equation}
{\nabla \cdot {\left(\frac{3-R_2}{2}{\bf v}E\right)}\over{\nabla \cdot \left(\frac{c \lambda}{\kappa_R \rho} \nabla E\right)}} \ll 1,
\end{equation}
where $R_2 = \lambda + {\lambda}^2 R^2$ is the Eddington factor.

Without radiative transfer, the energy exchange term in  (5) disappears, and we close the system of equations with
a barotropic EOS for the gas pressure:
\begin{equation}
P = \rho c_{\rm s} ^ 2 +  \left({{\rho} \over {\rho_{\rm c}}}\right)^ {\gamma} \rho_{\rm c} c_{\rm s}^2, 
\end{equation}
where  $c_{\rm s} = ({k_{\rm B} T }/{ \mu})^{1/2}$ is the isothermal sound speed, $\gamma=5/3$, the average molecular weight $\mu =2.33m_{\rm H}$, and the critical density, $\rho_{\rm c}=2\times 10^{-13}$ g cm$^{-3}$. The value of $\mu$ reflects an assumed gas composition of $n_{\rm He} = 0.1n_{\rm H}$.
The critical density determines the transition from isothermal to adiabatic regimes, and  we adopt a value to agree with the full angle-dependent 1D radiation-hydrodynamic calculation by \citet{masunaga98} that agrees with the collapse solution prior to H$_2$ dissociation.

We use the Truelove criterion to determine the addition of new AMR grids so that the gas density in the calculations always satisfies:
\begin{equation}
{\rho} < {\rho_{\rm J}} = {{J^2\pi c_{\rm s}^2}\over{G(\Delta x_l)^2}} \label{jeans},
\end{equation}
where $\Delta x_l$ is the cell size on level $l$, and we adopt a Jeans number of $J=0.25$ \citep{truelove97}.
In the case with radiative transfer, it is important to adequately resolve spatial gradients in the radiation field. Radiation gradients are primarily associated with collapsing regions hosting a star but are not covered by the Jeans gravitational criterion. We find that we adequately resolve the radiation field and avoid effects such as grid imprinting by refining wherever $\del E/E > 0.25$. Although the simulation box and gas behavior is periodic, we adopt Marshak boundary conditions for the radiation field.
This allows the radiation to escape from from the box as it would from a molecular cloud.

We insert sink, or star, particles 
in regions of the flow that have exceeded the Jeans density on the maximum level 
\citep{krumholz04}. These particles mark collapsing regions and also represent  protostellar objects.  In the simulation with radiative transfer, the particles act as radiative sources, and they are endowed with a sub-grid stellar model.  We describe the details of this model and its implementation in Appendix B.
By construction, stars that approach within eight cells are merged together. Small sink particles, such as those generated by disk fragmentation, tend to accrete little mass and frequently merge with their more substantial neighbors within a few orbital times.

\begin{figure*}
\epsscale{0.85}
\plotone{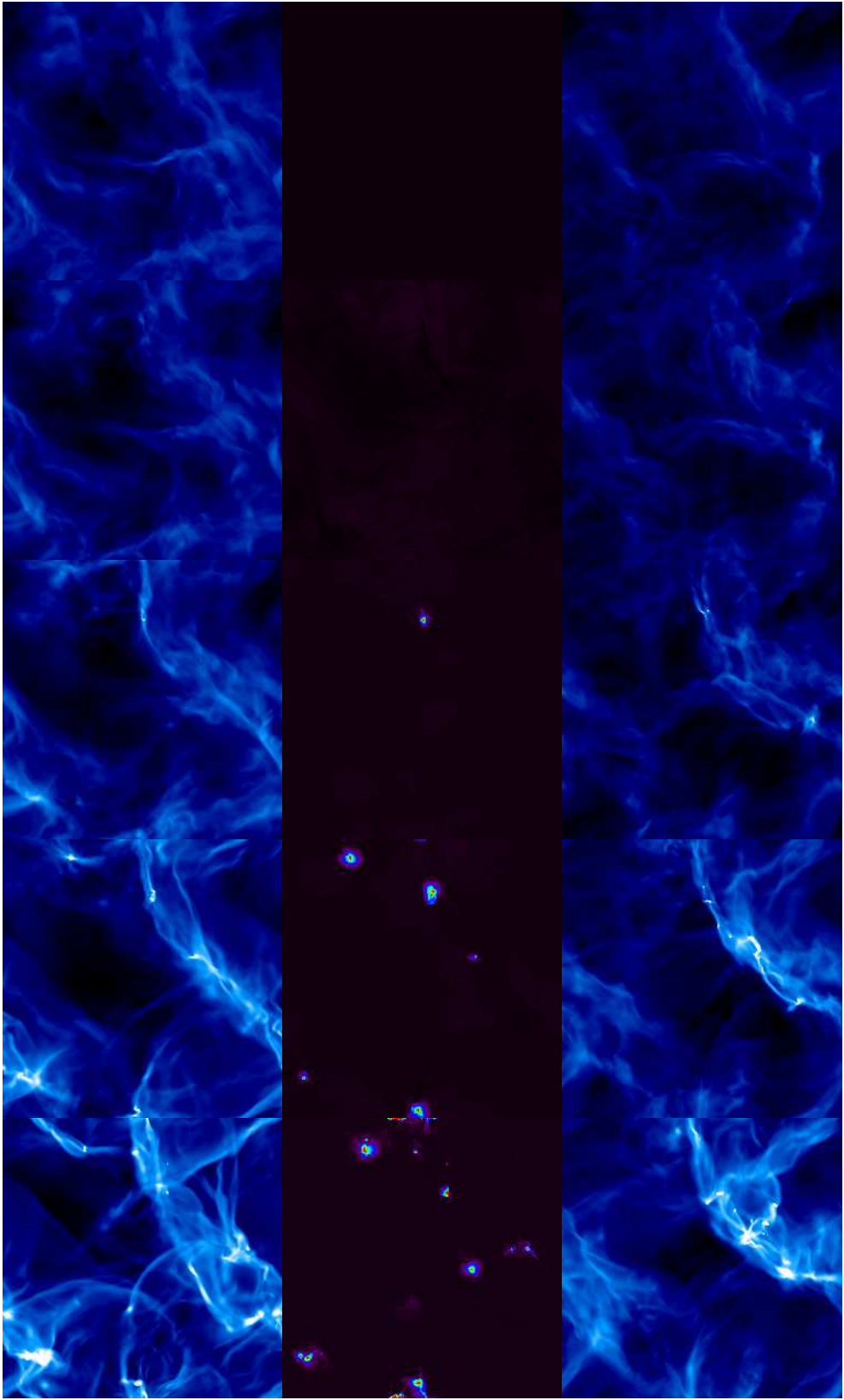} 
\epsscale{0.7}
\caption{Log gas column density of the RT (left) and NRT (right) simulations at 0.0, 0.25, 0.5, 0.75, 1.0 $t_{\rm ff}$. The log density weighted gas temperature for the RT is shown in the center. The color scale for the column density ranges from $10^{-1.5}-10^{0.5}$ g cm$^{-2}$ and $10-50$ K for the gas temperature. 
Animations of the left and right panels, as well as color figures, are included in the online version.
\label{isocolumn}}
\end{figure*}

\subsection{Initial Conditions}

We chose a characteristic
3D turbulent Mach number, ${{\mathcal M}}$=6.6, and assume that the cloud
is approximately virialized:
\begin{equation}
{ \alpha_{\rm vir}} = {{5 \sigma^2} \over { G M / R}} \simeq 1. 
 \end{equation}
The initial box temperature is $T$ = 10 K, length of the box $L$ = 0.65 pc and the average density is $\rho = 4.46 \times 10^{-20}$ g cm$^{-3}$, so that the cloud satisfies the observed linewidth-size relation \citep{solomon87, heyer04}. The total box mass is 185 $\msun$.

To obtain the initial turbulent conditions, we begin without self-gravity and apply velocity perturbations to an initially constant density field using the method described 
in \citet{maclow99}. These perturbations correspond to a 
Gaussian random field with flat power spectrum in the range $1 \le k \le 2$ where 
$k=k_{\rm phys} L / 2 \pi$ is the normalized wavenumber. 
At the end of three cloud crossing times, the turbulence follows a Burgers power 
spectrum, $P(k) \propto k^{-2}$, as expected for hydrodynamic systems of supersonic shocks. 
We denote this time $t = 0$. We then
turn on gravity and follow the
subsequent gravitational collapse for one global freefall time:
\begin{equation}
{\bar t}_{ff}= \sqrt{ {{3 \pi} \over {32 G \bar{ \rho}}}} = 0.315 \mbox{ Myr},
\end{equation} 
where $\bar \rho$ is the mean box density.
We continue turbulent driving in the simulations, using a constant energy injection rate to ensure that the turbulence does not decay and the cloud maintains approximate energy equipartition.

\begin{figure*}
\epsscale{1.1}
\plottwo{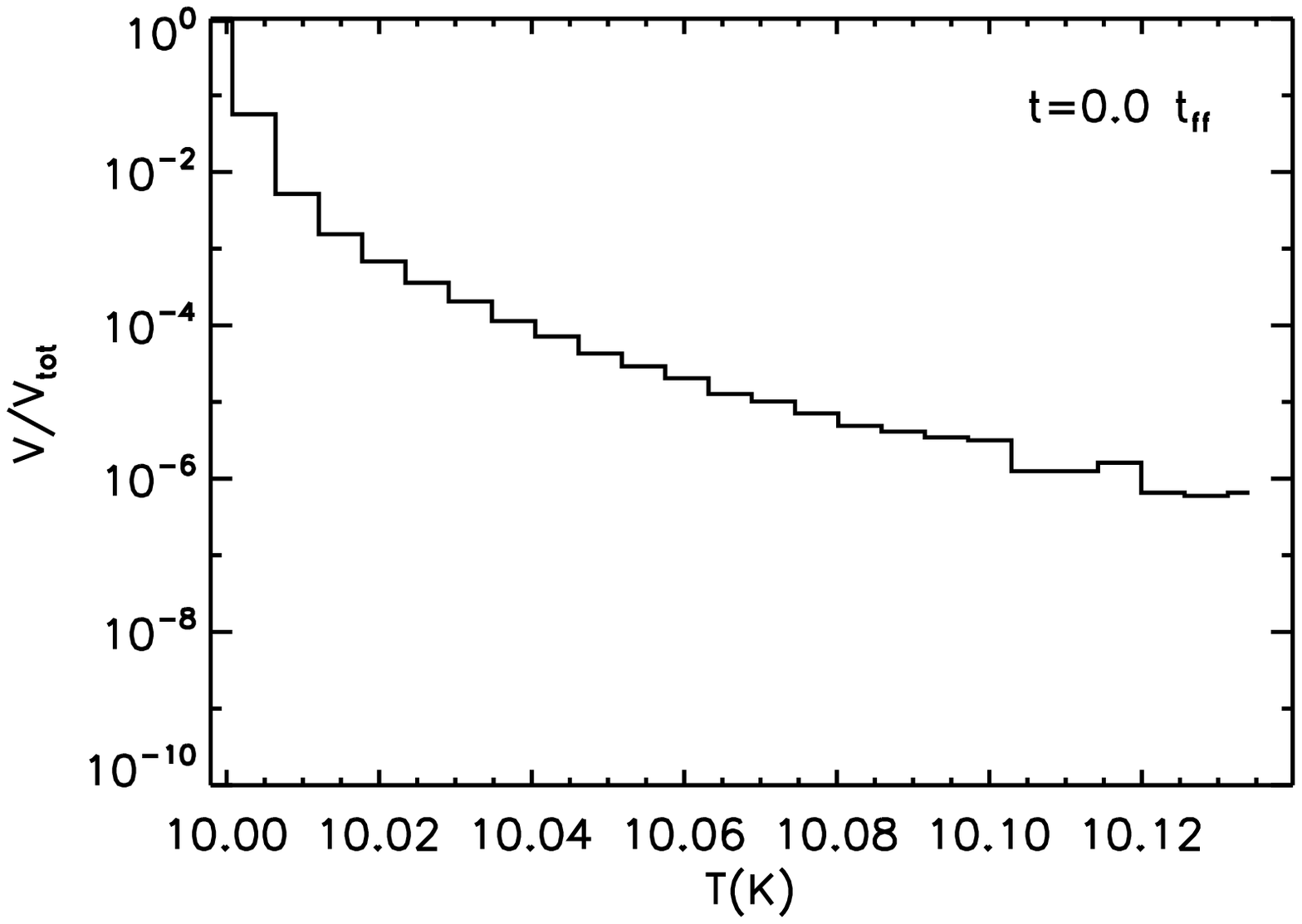}{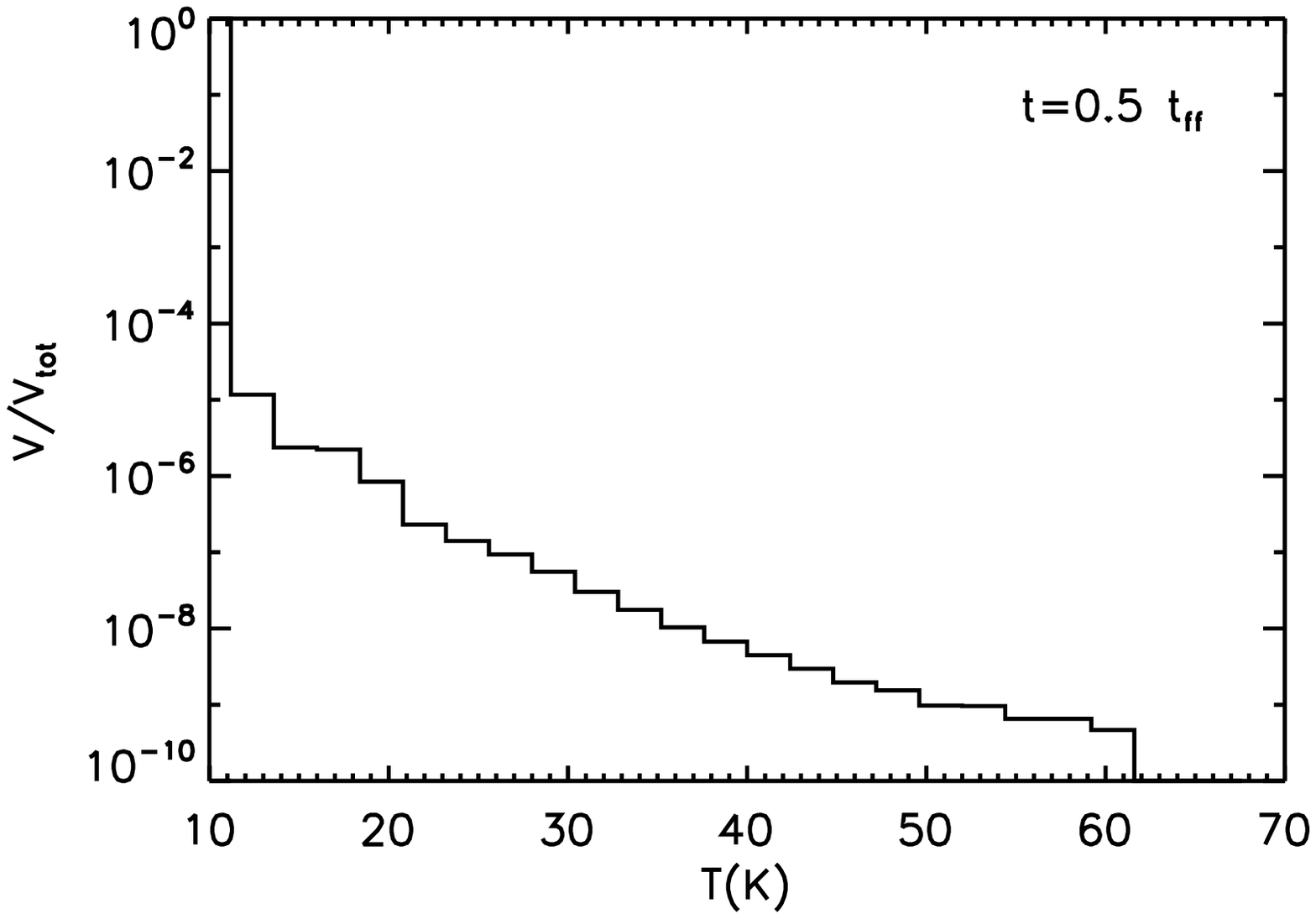}
\plottwo{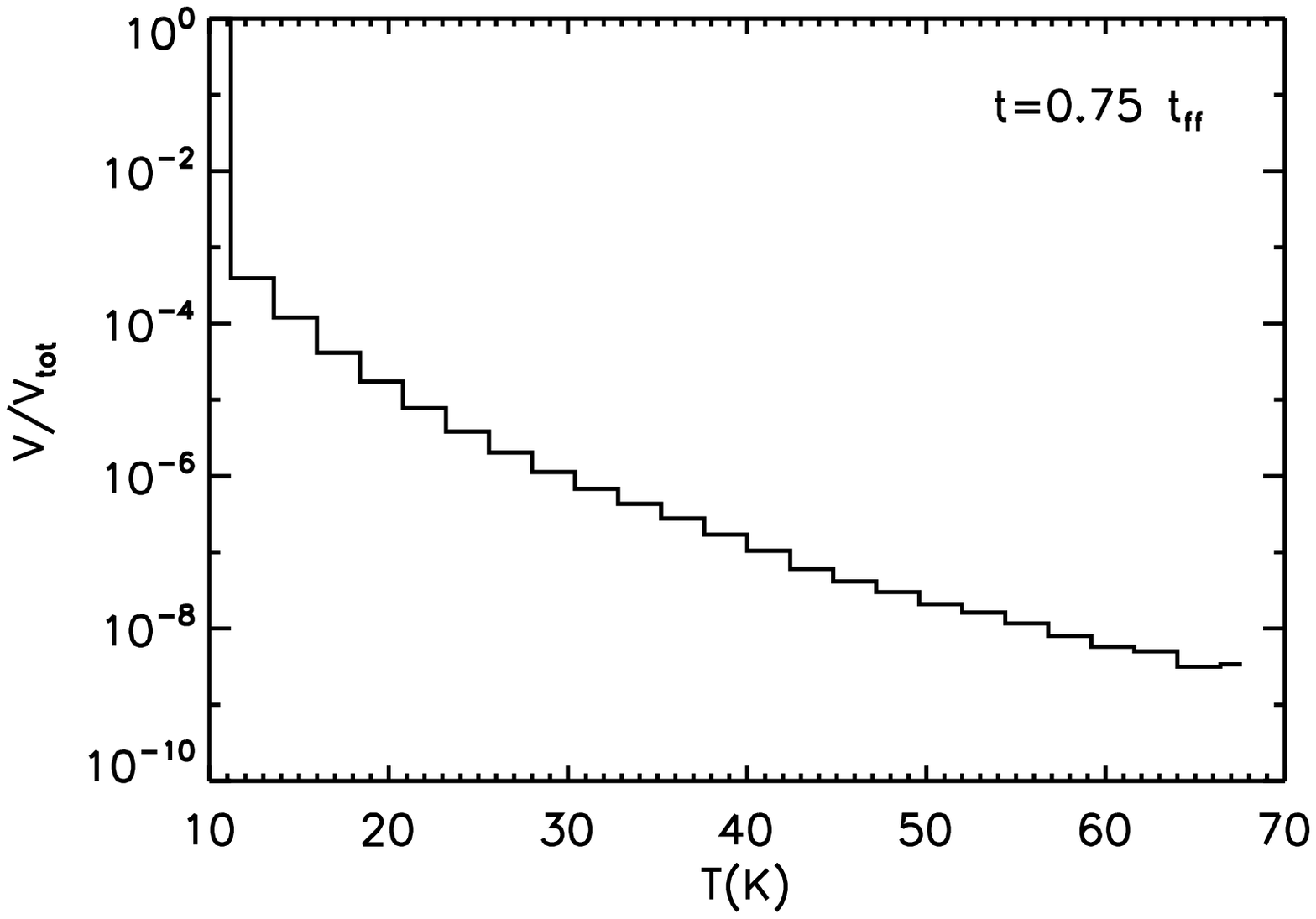}{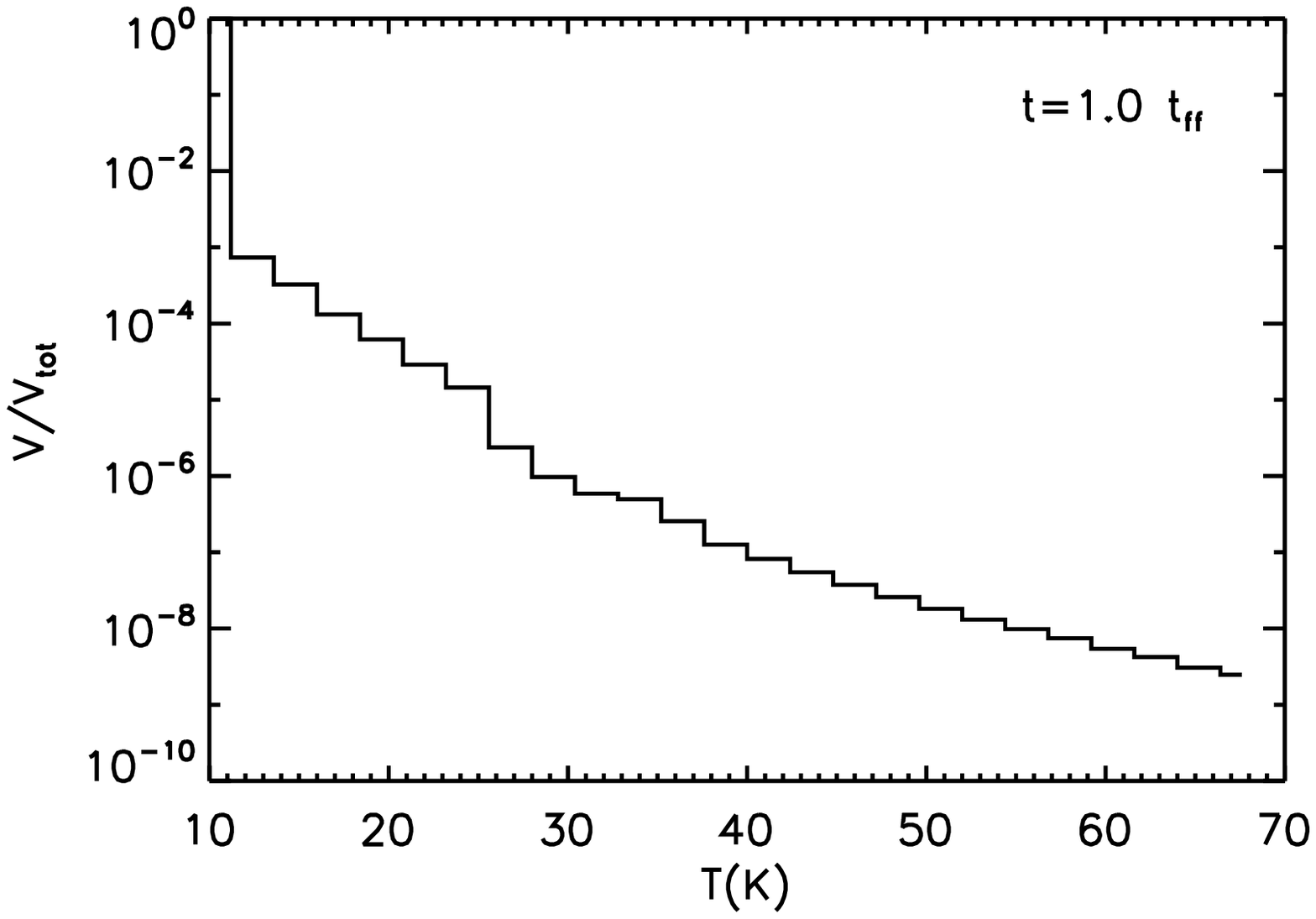}
\caption{Histogram of the gas temperatures weighted by volume fraction for RT at 0.0, 0.5, 0.75, and 1.0 $t_{ff}$. 
\label{temphist}}
\end{figure*}

The calculations have a $256^3$ base grid with 4 levels of factors of 2
in grid refinement, giving an effective resolution of $4096^3$, where $\Delta x_4$  =  32 AU. 
In \S 3.2, we describe the results of a high-resolution core study using 7 levels of 
refinement for an effective resolution of $65,536^3$ and minimum cell 
size  $\Delta x_7$ = 4 AU. 
Generally, the calculations run on 128-256 CPUs.

\begin{deluxetable*}{lcccccc}
\tablecaption{RT protostar properties at 1 $t_{\rm ff}$ \label{table1}}
\tablehead{\colhead{M (\msun)} & \colhead{$\dot M$ (\msun yr$^{-1}$)\tablenotemark{a}} &
\colhead{$\dot M_f$ (\msun yr$^{-1}$)\tablenotemark{b}} &
\colhead{$\bar{\dot M}$ (\msun yr$^{-1}$)\tablenotemark{c}} &
\colhead{L (\lsun)} & \colhead{Age (Myr) \tablenotemark{d}}}
\startdata
1.52 & $4.2\times 10^{-9}$ &  $1.1\times 10^{-7}$ & $8.7\times 10^{-6}$ &7.2 & 0.18 \\
0.45 & $2.0 \times 10^{-8}$ & $3.9\times 10^{-8}$ & $4.0\times 10^{-6}$ & 0.9&  0.11\\
0.09 & $1.4\times 10^{-7}$ &  $1.3\times 10^{-7}$ & $8.0\times 10^{-7}$ &0.3 &  0.11\\
2.91 & $8.1\times 10^{-7}$ &  $1.7\times 10^{-5}$& $2.9\times 10^{-5}$ & 177.5 & 0.10\\
0.35 & $5.6\times 10^{-7}$ & $2.0\times 10^{-7}$& $3.5\times 10^{-6}$& 1.3 & 0.10 \\
2.21 & $6.0\times 10^{-7}$ & $4.2\times 10^{-6}$& $2.4\times 10^{-5}$& 45.2 & 0.09\\
1.54 & $4.0\times 10^{-6}$ & $7.5\times 10^{-6}$& $1.7\times 10^{-6}$& 74.6 & 0.09 \\
1.17 & $9.8\times 10^{-6}$ & $1.7\times 10^{-5}$& $1.4\times 10^{-5}$& 69.4 & 0.09 \\
0.43 & $1.2\times 10^{-6}$ & $2.8\times 10^{-6}$& $6.0\times 10^{-6}$& 8.6 & 0.09\\
0.48 & $3.2\times 10^{-6}$ & $7.2\times 10^{-6}$&  $6.9\times 10^{-6}$ &19.4 & 0.08 \\
0.65 & $1.6\times 10^{-6}$ & $9.9\times 10^{-6}$& $1.1\times 10^{-5}$& 12.9 & 0.08\\
0.80 & $5.7\times 10^{-6}$ & $1.7\times 10^{-5}$& $1.5\times 10^{-5}$& 67.6 & 0.06 \\
0.33 & $2.1\times 10^{-5}$ & $2.2\times 10^{-5}$& $2.3\times 10^{-5}$& 79.1 & 0.02\\
0.06 & $4.7\times 10^{-6}$ & $5.1\times 10^{-6}$& $7.4\times 10^{-6}$& 3.9 & 0.01\\
0.01 & $3.0\times 10^{-6}$ & $1.1\times 10^{-5}$& $8.6\times 10^{-6}$& 0.8 & 0.003\\
\enddata
\tablenotetext{a} {Instantaneous accretion rate.}
\tablenotetext{b} {Final accretion rate averaged over the last $\sim$ 2500 yrs.}
\tablenotetext{c} {Mean accretion rate averaged over the protostar lifetime.}
\tablenotetext{d} {Age calculated from the time of particle formation.}
\end{deluxetable*}

\section{RESULTS \label{results}}

\subsection{Radiative Transfer and Non-Radiative Transfer Comparison}

In order to quantify the effects of radiative feedback on low-mass star formation, we compare a  
simulation including radiative transfer with a non-radiative one using an EOS.  
The latter simulation is essentially isothermal throughout since the highest density allowed by the Truelove criteria at the fiducial maximum AMR level corresponds to $\rho \simeq 5 \times 10^{-15}$ g cm$^{-3}$. 
With the adopted EOS, gas of this density is not heated above 11 K.

Images of the two simulations at different times are shown in Figure \ref{isocolumn}. Although the simulations use identical forcing patterns applied at the same Mach number,  the details of the turbulence differ as the two calculations have slightly different time steps and turbulent decay rates. 
Both calculations begin at $t=0$ with a centrally condensed region that forms the first stars, an imprint of the large wavenumber driving.
Once gravity is turned on, we continue driving the simulations with the same energy injection rate, yielding  3D Mach numbers of  7.0 and 8.6 at 1 $t_{\rm ff}$ for the NRT and RT calculations, respectively. 
Because gravitational collapse causes non-turbulent velocity motions, we chose to fix the energy injection rate rather than the total kinetic energy. Thus, the root-mean-squared gas velocity no longer exactly indicates the total turbulent energy, and the Mach number increases above the initial value.
In Tables \ref{table1} and \ref{table2}, we list the properties of the stars formed in each calculation at 1$t_{\rm ff}$.

\subsubsection{Temperature Distribution}

At $t=0$, the RT simulation is nearly isothermal and gas temperatures, are distributed between 10-11 K (Figure \ref{temphist}). Evaluated at the mean box density, the gas is quite optically thin with an average optical depth though the box of $\tau_{\rm L} = L \times \kappa_{\rm R} \rho = 0.65~\mbox{pc} \times 4.46\times10^{-20}$ g cm$^{-3} \times 0.2~\mbox{cm}^2~\mbox{g}^{-1}  \sim 0.02$. Since the box is so transparent, the gas cools very efficiently. Small temperature variations arise in the initial state due to the distribution of strong shocks. 
For reference, gas compressed by a Mach 10 shock at 10 K will undergo net heating of $<$ 0.1 K during a time step. Qualitatively, the change is so small because the radiative cooling time is a factor of $\sim 10^3$ smaller than the time step.

\begin{figure}
\epsscale{1.2}
\plotone{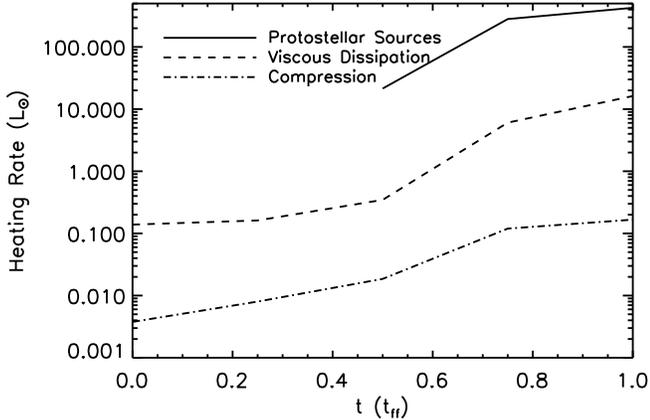}
\caption{The magnitude of the heating rate due to all stellar sources, viscous dissipation, and gas compression at the times shown in Figure \ref{isocolumn}. \label{heating}}
\end{figure}

Under the influence of gravity, collapsing regions begin to become optically thick, where individual cells at the maximum simulation densities reach optical depths of $\tau \simeq 3$ when $T=100$ K.  Figure \ref{temphist} shows the evolution of the gas temperature distribution over a freefall time. 
There are three processes that result in heating. First, there is the direct contribution from the protostars, which we add as a source term in the radiation energy equation. Second there is heating due to viscous dissipation, which is given by:
\begin{equation}
\dot e_{\rm vis} = -(\sigma ' \cdot \del) \cdot {\bf v},
\end{equation}
where  $\sigma '$ is the viscous stress tensor, $\sigma ' = \eta(S -{2 \over 3} I \del \cdot v)$ and $S = \del v + (\del v)^T$ \citep{landau}. 
 We assume that the dynamic viscosity $\eta= \rho |{\bf v}| \Delta x / { \mathcal R}e_g$, where the Reynolds number,  ${\mathcal R }e_g \simeq 1$,  
at the dissipation scale. However, turbulent dissipation occurs over a range of the smallest scales on the domain, where the largest amount of dissipation occurs on the size scale of a cell.
Thus, we expect this formula to be uncertain to within a factor of two. Third, the net heating due to gas compression is given by:
\begin{equation}
\dot e_{\rm comp} = - P( \del \cdot  {\bf v} );
\end{equation}
the heating is negative (i.e., cooling occurs) in rarefactions.
Figure \ref{heating} shows the heating contributions summed over the entire domain. At $t=0$, the only source of heating is turbulent motions. The figure demonstrates that after star formation commences protostellar output rather than compression is responsible for the majority of the heated gas, and at $1t_{\rm ff}$ protostellar heating dominates by an order of magnitude relative to viscous dissipation and four orders of magnitude relative to gas compression.  Viscous dissipation dominates the heating prior to star formation. After star formation is underway, viscous dissipation occurs in the protostellar disks. In contrast, turbulent shocks then contribute very little to the total.

Figure \ref{temphist} shows the evolution of the gas temperature distribution over a freefall time. The amount of heated gas ($T > 12$ K) increases with the number of stellar sources from 0.06\% of the volume for one protostar at 0.5 $t_{\rm ff}$ to $\sim 4$\% at $1~t_{\rm ff}$. 
The corresponding mass fractions of the heated gas are slightly higher at 0.3 \% and 5\%, respectively.
 As we have seen in the previous figure, most of this heating is directly related to the protostars, and it comprises a relatively small volume filling fraction.
 
\begin{figure}
\epsscale{1.2}
\plotone{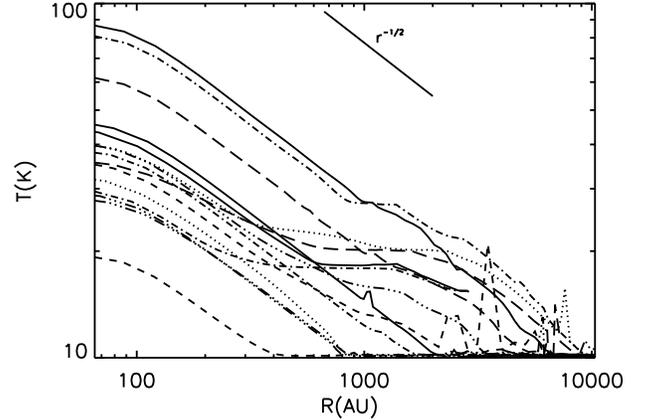}
\caption{The gas temperature as a function of distance from the source for all sources in the RT simulation at 1.0 $t_{ff}$.  The line indicates a line with $T \propto r^{-1/2}$. \label{tempvsr}}
\end{figure}

 The temperature distribution as a function of distance from the sources is shown in Figure \ref{tempvsr}.  As illustrated, heating is local and occurs within $\sim$ 0.05 pc of the protostar. Since the remainder of the cloud remains near 10 K, additional turbulent fragmentation is not affected by pre-existing protostars. However, radiative feedback profoundly influences the evolution of the protostars, accretion disks, and stellar multiplicity as we will show (see \S 3.1.2-3.1.3). Our temperature profiles are qualitatively similar to those of 
 \citet{masunaga00}, who model 1D protostellar collapse with radiative transfer.
During the formation of the low-mass protostar, \citet{masunaga00} also find that heating above 10 K is confined within 0.05 pc of the central source and that significant variation in temperature occurs as a function of density and time. 
Additional studies using GFLD \citep{whitehouse06} or approximate radiative transfer methods \citep{stamatellos07, forgan09} find similar heating beyond that expected from a barotropic EOS.

\begin{figure*}
\epsscale{1.15}
\plottwo{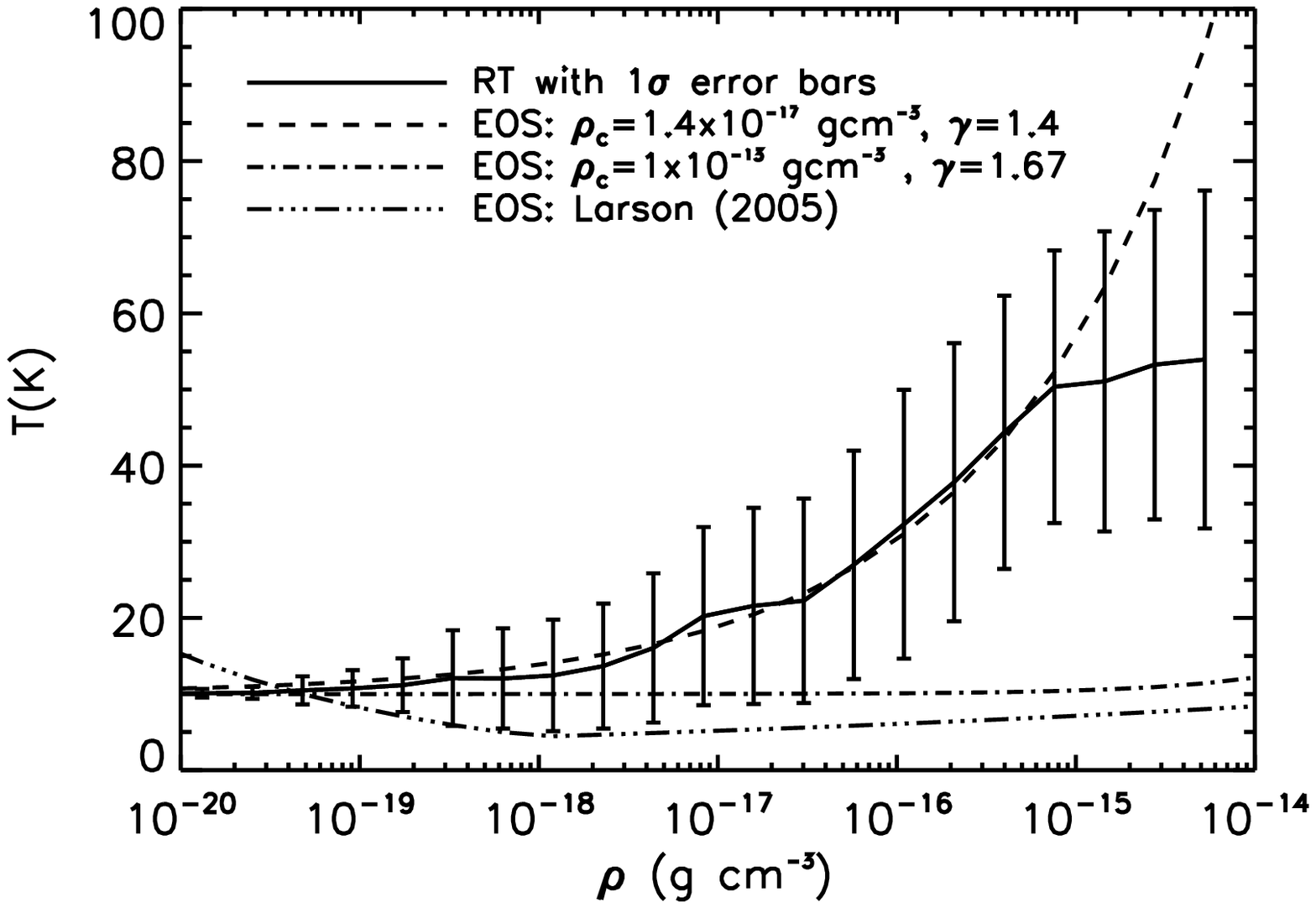}{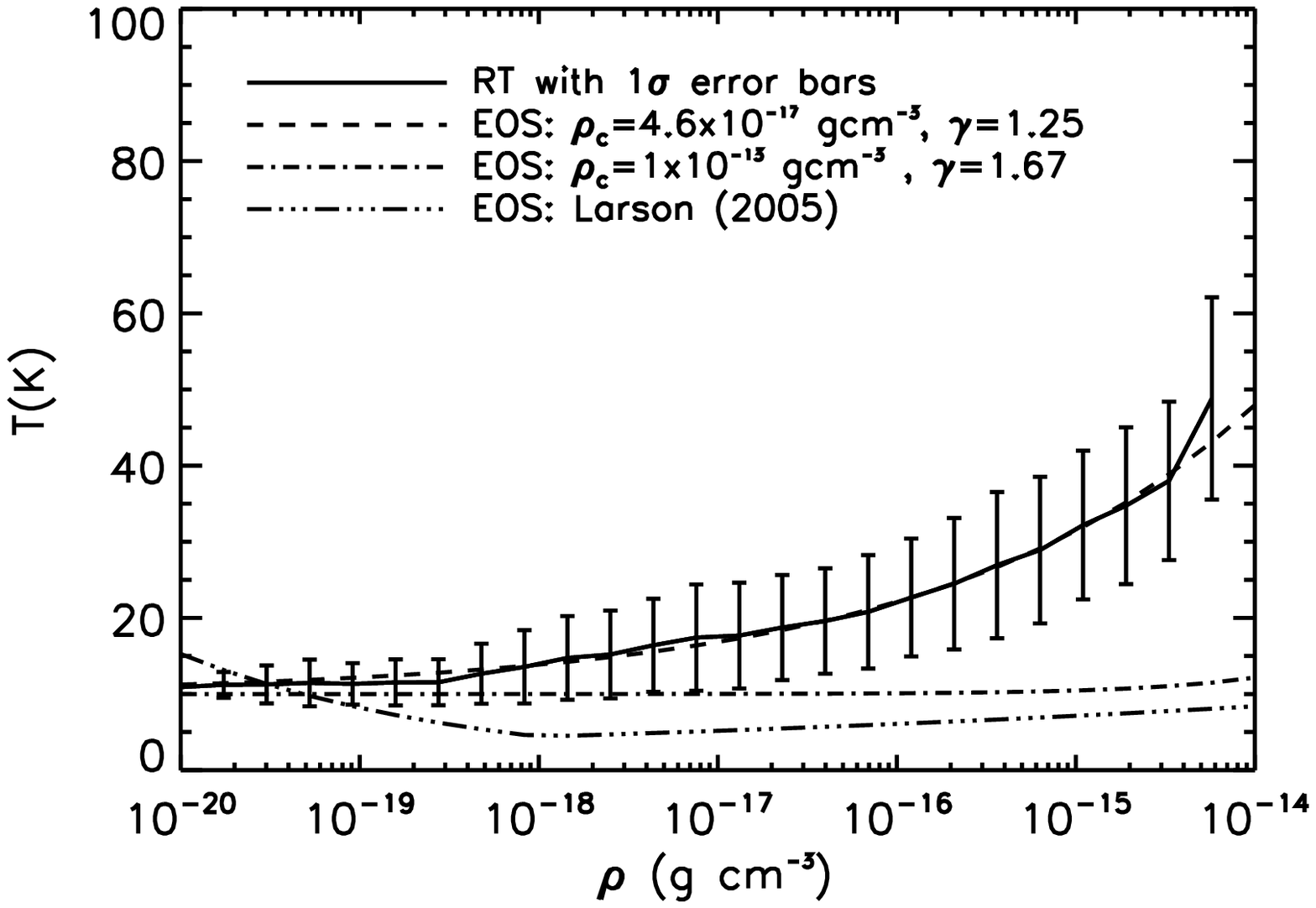}
\caption{The average gas temperature at 0.8 $t_{ff}$ and 1.0 $t_{ff}$ as a function of density. The error bars give the temperature dispersion in each bin. The dashed line is a least-sqares fit of  Equation 11 which returns $\rho_{\rm c}$  and $\gamma$. The dot-dashed line plots Equation 11 with the original parameters: $\rho_{\rm c} = 1 \times 10^{-13}$ g cm$^{-3}$ and $\gamma = 1.67$. The power law density-temperature relation from \citet{larson05} is also plotted.
\label{tvsrho}}
\end{figure*}

Due to temperature variation with both density and time, we find that the gas temperature is poorly represented by a single EOS with characteristic critical density and $\gamma$. Figure \ref{tvsrho} shows the distribution of cell temperatures as a function of cell density. For reference, we also plot our fiducial EOS for the NRT simulation as well as the EOS presented by \citet{larson05}. We find that many cells at lower densities are heated due to close proximity with a source. In fact, for both the EOS described in \S 2, which only includes the heating due to gas compression, and the \citet{larson05} EOS, none of the cells are predicted to heat much above the initial 10 K temperature.

Nonetheless, at any given time a representative EOS can be formulated by fitting 
the mean grid cell temperature binned  
as a function of density. Figure \ref{tvsrho} shows a least-squares fit of the temperature data for two different times.  The magnitude of the error bars is given by the standard deviation of the temperatures in each density bin.
Because such an equation fits the average temperature, there is necessarily a large scatter as illustrated by the error bars. The two fits return different effective critical densities and gamma values. Thus, a single EOS results in a large fraction of cells unavoidably at the wrong temperature.

\begin{deluxetable*}{lccccc}
\tablecaption{NRT protostar properties at 1 $t_{\rm ff}$ \label{table2}}
\tablehead{\colhead{M (\msun)} & \colhead{$\dot M$ (\msun yr$^{-1}$)\tablenotemark{a}}  &
\colhead{$\dot M_f$ (\msun yr$^{-1}$)\tablenotemark{b}} &
\colhead{$\bar{\dot M}$ (\msun yr$^{-1}$)\tablenotemark{c}} &
 \colhead{Age (Myr) \tablenotemark{d}}}
\startdata
3.92 & $7.2\times 10^{-6}$ & $1.2\times 10^{-5}$& $2.2\times 10^{-5}$& 0.15 \\
4.77 & $1.6 \times 10^{-6}$ & $2.7\times 10^{-5}$& $2.6\times 10^{-5}$&  0.15\\
2.91 & $1.0\times 10^{-5}$ & $1.2\times 10^{-5}$& $2.6\times 10^{-5}$&  0.11\\
4.84 & $2.1\times 10^{-5}$ & $2.3\times 10^{-5}$& $4.4\times 10^{-5}$& 0.11\\
0.66 & $2.5\times 10^{-7}$ & $2.7\times 10^{-7}$& $7.6\times 10^{-6}$& 0.09 \\
1.13 & $1.3\times 10^{-5}$ & $1.3\times 10^{-5}$& $2.1\times 10^{-5}$& 0.05\\
0.66 & $8.9\times 10^{-7}$ & $9.0\times 10^{-7}$& $1.2\times 10^{-5}$& 0.05 \\
0.55 & $9.3\times 10^{-7}$ & $1.0\times 10^{-6}$& $1.1\times 10^{-5}$& 0.05 \\
0.71 & $5.9\times 10^{-6}$ & $5.6\times 10^{-6}$& $1.4\times 10^{-5}$& 0.05\\
1.32 & $1.2\times 10^{-5}$ & $1.4\times 10^{-5}$& $7.8\times 10^{-5}$& 0.02 \\
0.08 & $2.7\times 10^{-5}$ & $6.6\times 10^{-6}$& $5.9\times 10^{-6}$& 0.02\\
0.49 & $1.1\times 10^{-5}$ & $1.1\times 10^{-5}$& $3.6\times 10^{-5}$& 0.02 \\
0.26 & $5.0\times 10^{-6}$ & $1.2\times 10^{-5}$& $2.0\times 10^{-5}$& 0.02\\
0.04 & $5.8\times 10^{-6}$ & $1.1\times 10^{-5}$& $2.8\times 10^{-6}$& 0.02\\
0.02 & $2.6\times 10^{-9}$ & $1.1\times 10^{-8}$& $1.2\times 10^{-6}$& 0.02\\
0.12 & $1.3\times 10^{-6}$ & $1.5\times 10^{-6}$& $9.3\times 10^{-6}$& 0.02\\
0.04 & $2.7\times 10^{-6}$ & $2.7\times 10^{-6}$& $4.2\times 10^{-6}$& 0.01\\
0.01 & $8.6\times 10^{-12}$ & $5.5\times 10^{-12}$& $1.8\times 10^{-6}$& 0.01\\
0.09 & $6.4\times 10^{-6}$ & $5.1\times 10^{-6}$& $1.3\times 10^{-5}$& 0.01\\
0.14 & $1.9\times 10^{-5}$ & $1.9\times 10^{-5}$& $2.3\times 10^{-5}$& 0.01\\
0.02 & $2.7\times 10^{-6}$ & $7.0\times 10^{-6}$& $5.5\times 10^{-6}$& 0.01\\
0.05 & $2.4\times 10^{-5}$ & $1.5\times 10^{-5}$& $1.6\times 10^{-5}$& 0.01\\
\enddata
\tablenotetext{a} {Instantaneous accretion rate.}
\tablenotetext{b} {Final accretion rate averaged over the last $\sim$ 2500 yrs.}
\tablenotetext{c} {Mean accretion rate averaged over the protostar lifetime.}
\tablenotetext{d} {Age calculated from the time of particle formation.}
\end{deluxetable*}

Since accretion luminosity is predominantly emitted at the stellar surface, a low simulation resolution, when not augmented for the missing source contribution, can significantly neglect a large part of the heating (e.g., \citealt{bate09b}). 
Typical pre-main sequence protostellar radii are expected to range from 3-5 $R_{\odot}$ for low-mass stars \citep{palla93, robit06}.
Thus, the temperature at a distance, $r$, from an emitting source, $L_*$, is given by:
\begin{equation}
 T = \left({L_* \over {4 \pi \sigma_B r^2}}\right)^{1/4},
\end{equation}
where $\sigma_{\rm B}$ is the Stefan-Boltzman constant, and the gas distribution is assumed to be spherically symmetric. Then the difference in accretion luminosity for a simulation with minimum resolution of $R_{\rm res} =$ 0.5 AU  versus a simulation resolving down to the stellar surface at $R_*=5~R_{\odot}$ is given by:
\begin{equation}
\Delta L = {{G m \dot m} \over {R_{res}}} \times \left( {{R_{\rm res}}\over{R_*}} - 1\right) \simeq {{G m \dot m} \over {R_{\rm res}}} \times (20) \label{deltal}.
\end{equation}
Thus, the actual accretion luminosity at the higher resolution is a factor of 20 larger! Since we adopt a stellar model to calculate the protostellar radii self-consistently, we include the entire accretion luminosity contribution down to the stellar surface in our simulations.
From  (\ref{deltal}), the difference in luminosity corresponds to a factor of $(20)^{1/4}$ or $ \sim 2$ underestimation of the gas temperature.  Nonetheless, this estimate is conservative since it does not include the additional luminosity emitted by the protostar, which may become significant during the Class  II and late Class I phase.
Thus, we expect the simulation of \citet{bate09b} may
overestimate the extent of small scale fragmentation and BDs formed in disks.

\subsubsection{Stellar Mass Distribution}

\begin{figure}
\epsscale{1.2}
\plotone{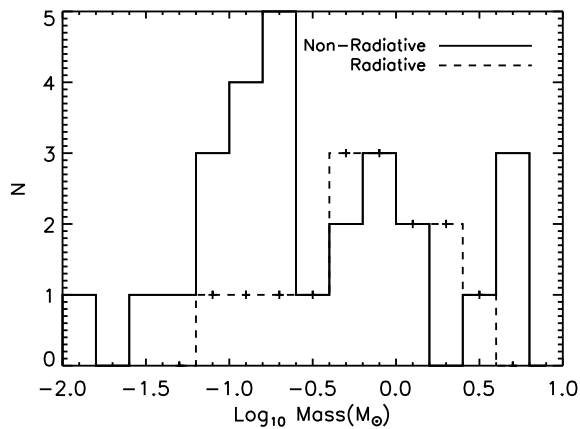}
\caption{The distribution of masses (star + disk) for the two simulations at 1.0 $t_{ff}$. The solid and dashed-cross lines indicate the NRT and RT simualtions, respectively.   \label{imf}}
\end{figure}

The large temperature range in the RT simulation has a profound effect on the stellar mass distribution. Figure \ref{imf} depicts the total mass of the star-disk systems in each simulation, were we define the surrounding disk as cells with $\rho > 5 \times 10^{-17}$ g cm$^{-3}$.
We find that this cutoff selects gas within a few hundred AU of the protostars, visually identified with the disk, while excluding the envelope gas.
Increased thermal support in the protostellar disk acts to suppress disk instability and secondary fragmentation in the core. In contrast, protostellar disks in the NRT calculation suffer high rates of fragmentation. Most of these small fragments are almost immediately accreted by the central protostar, driving temporarily large accretion rates onto the central source. 
If we define the star formation rate per freefall time as
\begin{equation}
\mbox{SFR}_{ff} = {\dot M_* \over {M/{\bar t_{ff}}}},
\end{equation}
\citep{krumholzmckee05} then the total star formation rate in the NRT simulation is 13\% versus 7\% in the RT simulation. Thus, the RT SFR$_{\rm ff}$ is almost half the NRT value and agrees better with observations \citep{KandT07}.
Since the simulations have the same numerical resolution, thermal physics must be directly responsible. In the RT simulation, cores containing protostars experience radiative feedback that slows collapse and accretion.

Due to the small number statistics, we do not directly compare with the shape of the observed initial mass function. Accurate comparison is also problematic because many of the late forming protostars are still actively accreting. As shown in Table \ref{table1}, by 1$t_{\rm ff}$ in the RT case, about a third, or 5 of the protostars, have accretion rates that are at least 5 times smaller than their individual mean accretion rate, indicating that the main accretion phase has ended. Adopting an efficiency factor of $\epsilon_{\rm core} = {{1}\over{3}}$ to account for mass loss due to outflows \citep{matzner00,alves07, enoch08}, we find that the mean protostellar mass of these protostars is $\bar m = 0.4$ \msun, which is comparable to the expected mean mass of the system initial mass function of $\sim$ 0.5 \msun \citep{scalo86, chabrier05}.

The dynamics of close bodies and embedding gas are difficult to accurately resolve inside a small number of grid cells, so we merge particles that pass within 8 cells. Without this limit, some of the small fragments would dynamically interact with the central body and be ejected from the stellar system. These brown dwarf size objects are commonly produced in simulations that do not include a merger criterion, typically in larger numbers than are observed in the stellar IMF (e.g., \citealt{bate03, bate05, bate09a}). As a result, the simulation IMF only resolves wide binaries with separations $>$ 300 AU.

\begin{figure*}
\epsscale{1.15}
\plottwo{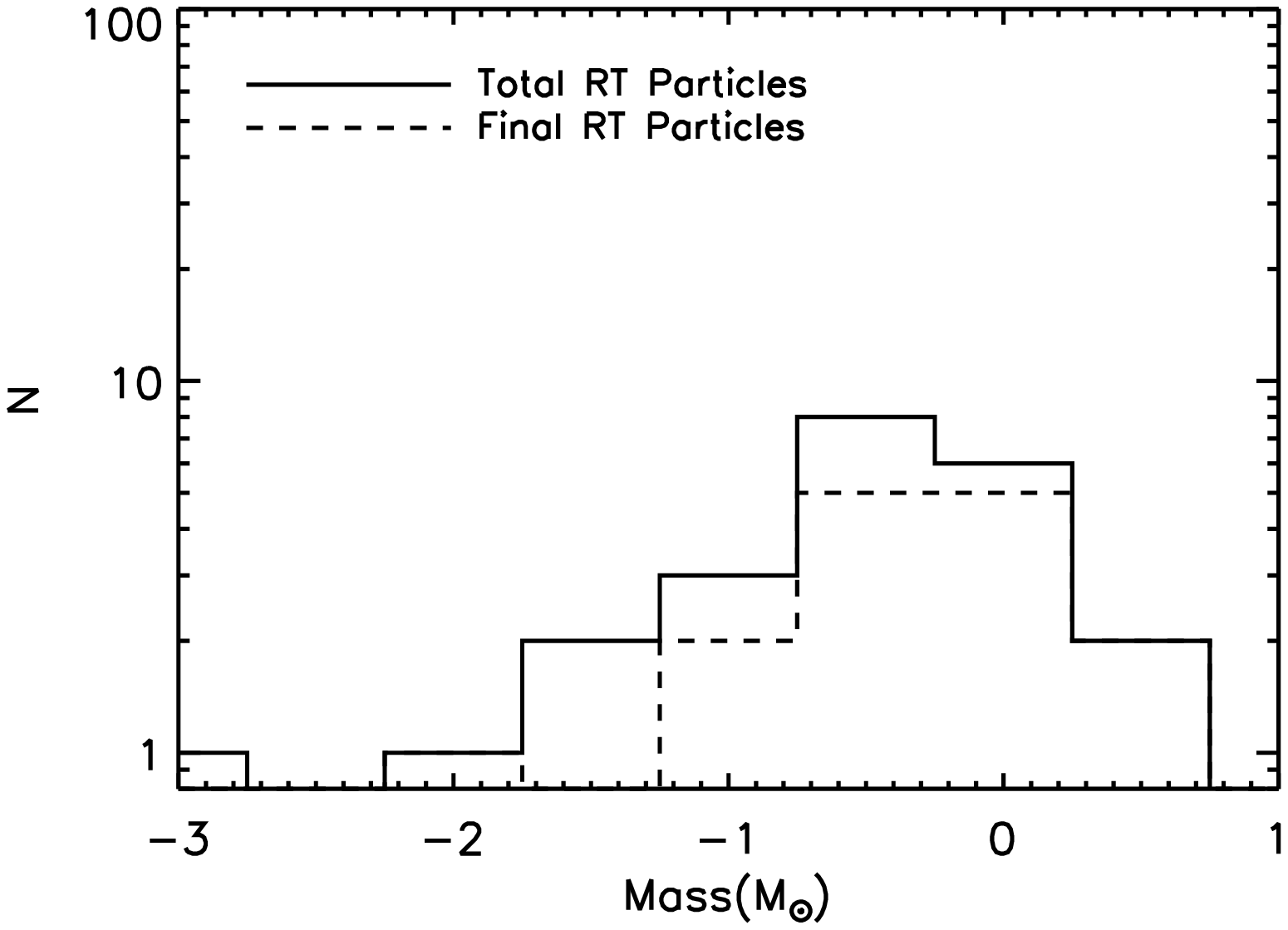}{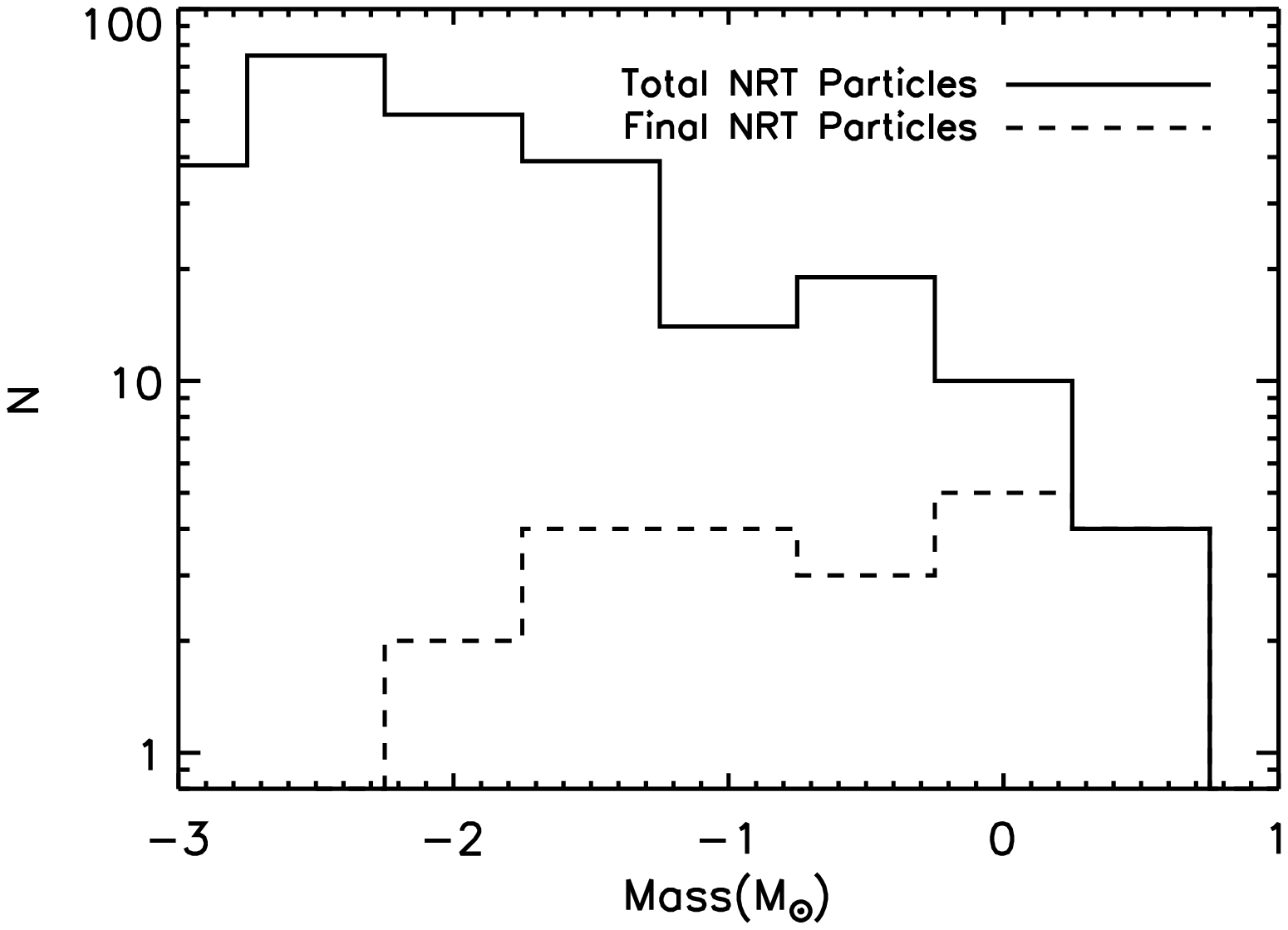}
\caption{The figures show the distribution of particles formed as a function of mass for the RT(left) and NRT(right) simulations (solid line). These include the particles that are merged, where the total particle number with final masses greater than $10^{-3}$ \msun is 23 and 251, respectively. The dashed lines show the distributions of stellar masses at the final time output. \label{frag}}
\end{figure*}

Figure \ref{frag} shows a histogram of all created fragments in both simulations, including the final mass of the objects that are merged. Due to the low-resolution of the disks in the simulations, $\sim 10$ cells, the many small bodies shown in the NRT distribution indicate numerical disk instability rather than small bodies forming from gravitational collapse. 
The large number of particles that are created in the NRT case is directly related to the nearly isothermal EOS.  Gravitational
instability in disks results in filamentary spiral arms. If the gas is isothermal, the filaments
undergo indefinite collapse irrespective of the numerical resolution \citep{truelove97, truelove98, larson05, inutsuka92}.
In a numerical calculation, this means that all the cells along a filament exceed the Jeans criterion nearly simultaneously and trigger refinement. Once the maximum refinement level is reached, sink particles are introduced in cells with densities violating the Truelove criterion \citep{truelove97}.
Since our sink particle algorithm is formulated to represent a collapsing sphere, it is not well suited to filament collapse. Kratter et al.~(2009, in preparation) have addressed this issue in their predominantly isothermal simulations by transitioning from an isothermal to an adiabatic EOS once the density reaches a factor of four below the density at which sink particles are created. This has the effect of forcing filaments to fragment into quasi-spherical blobs prior to sink particle creation, thereby allowing the collapsing objects to be faithfully represented by point-like sink particles. 
At higher resolution, the barotropic nature of our EOS is invoked and so much of this fragmentation disappears (see Section \ref{convergenceEOS}).
Similarly, in radiative calculations filamentary collapse is halted by heating due to radiative feedback, so that fragmentation is described by spherical rather than filamentary collapse. For either representation of heating, although numerical fragmentation in filaments is restricted, physical fragmentation may yet occur.

The creation and fragmentation of filaments in the simulations is a result of gravitational instability driven by rapid accretion. The criterion for the onset of instability is similar to the classic Toomre $Q<1$ condition, slightly modified by the non-axisymmetry of the instabilities and the finite scale height of the disks, which is a result of turbulence driven by the accretion. This sort of gravitational instability has been investigated by Kratter et al.~(2008, 2009 in preparation), who point out that the presence or absence of instability depends largely on the accretion rate onto the disk. The rate of mass transport through an $\alpha$ disk is
\begin{equation}
\dot{m} = 3\left(\frac{\alpha}{Q}\right) \frac{c_{\rm s,disk}^3}{G},
\end{equation}
where $Q$ is the Toomre parameter for the disk and $c_{\rm s,disk}$ is the sound speed within it. Gravitational instabilities produce a maximum effective viscosity $\alpha\sim 1$. At early times, we find that the accretion rate from a core onto the disk forming within it can be $\gg c_{\rm s,core}^3 / G$, where $c_{\rm s,core}$ is the sound speed in the core. If the sound speeds in the disk and core are comparable, $c_{\rm s,disk} \sim c_{\rm s,core}$, as is the case in the low-resolution NRT simulation, then the disk can only deliver matter to the star at a rate $\sim c_{\rm s,core}^3 / G$ while still maintaining $Q > 1$. As a result matter falls onto the disk faster than the disk can deliver it to the star, and the disk mass grows, driving $Q$ toward 1 and producing instability and fragmentation, as illustrated by the NRT simulation. Conversely, if the disk is warmed, either by radiation or by a switch from an isothermal to an adiabatic equation of state, then $c_{\rm s,disk} > c_{\rm s,core}$ and the rate at which the disk can deliver gas to the star increases. If the disk is sufficiently warm then it can process all the incoming material while still maintaining $Q > 1$. As a result the disk does not fragment, as is seen in both the low- and high-resolution RT simulations and in high-resolution NRT simulation. This shows that the fragmentation in the low-resolution NRT simulation is indeed numerical rather than physical in origin, and that it is a result of the density-dependence of the equation of state rather than of the resolution directly.

This analysis also sheds light on the importance of numerical viscosity. \citet{krumholz04} show that in the inner few cells of disks, numerical viscosity can cause angular momentum transport at rates that correspond to $\alpha \ga 1$. However, as the analysis above shows, increasing $\alpha$ tends to suppress fragmentation rather than enhance it. We find that fragmentation is more prevalent in the low-resolution NRT simulation than the high-resolution one, which is exactly the opposite of what we would expect if numerical angular momentum transport were significantly influencing fragmentation. Therefore we conclude that numerical angular momentum transport is not dominant in determining when fragmentation occurs in our simulations.

In isothermal calculations, the issue of filamentary collapse is a problem for all sink particle methods and it is not unique to grid-based codes. Due to the filamentary fragmentation in the NRT case, we prefer to merge close particle pairs in the simulations rather than follow their trajectories. 
Note that particles are inserted with the mass exceeding the Truelove criterion rather than the net unstable mass in the violating cells. Particles created within a discrete bound mass typically gain size quickly. 
Most particles formed in the unstable disk regions form in a spiral filament and do not have significant bound mass, so the particle mass is tiny when they are accreted by the central object. 
However, if several small particles are created within the merging radius each time step around a particular protostar, their merging can significantly increase the instantaneous accretion rate.
As illustrated by the figure, there are only a handful of objects that form and approach within a merging radius in the RT simulation, whereas the NRT simulation produces a plethora of such bodies.  

\citet{bate09b} finds a similar reduction in protostar number with the addition of radiative transfer. As in our calculation, the final number of stars including radiative transfer is sufficiently small that a statistical comparison with the IMF is problematic. Instead, we base our comparison on the mean stellar mass. Using a resolution of 0.5 AU, \citet{bate09b} finds $\bar m \sim 0.5 ~\msun$, which does not include outflows or any scaling factor accounting for their presence. Adopting a scaling factor of $\epsilon_{\rm core} = 1/3$ would produce a mean of $\bar m \sim  0.2~ \msun$, lower than our RT mean mass and the mean mass of  either the system or individual stellar initial mass function reported by \citet{chabrier05}. However, in \citet{bate09b} a number of the protostars are continuing to accrete and have not reached their final mass. In addition,  \citet{bate09b} demonstrates that the mean stellar mass increases as calculations approach higher resolution and include a larger portion of the accretion luminosity. This result is most likely because disk fragmentation decreases as the gas becomes hotter, thus increasing accretion onto primary objects.  It is possible that if \citet{bate09b} had included all the accretion and stellar luminosity, the mean mass obtained would be closer to the value we find.
%

Observations suggest that BDs compose $\sim 30$\% of the total population of clusters \citep{andersen06}. 
Despite the merger criterion we adopt, the NRT calculation produces a significant number of BDs, $> 30 $\% sans scaling with $\epsilon_{\rm core}$, resulting in a slightly lower mean mass than the RT run.  In comparison, \citet{bate03} find that approximately half of the objects formed are BDs, resulting in a mean mass of $\sim 0.1 ~\msun$. This result persists for barotropic calculations modeling more massive clusters with superior resolution \citep{bate09a}. Calculations using a modified EOS that includes effects due to the internal energy and dissociation of H$_2$,  ionization state of H, and approximate dust cooling find increased disk fragmentation, leading to numerous BDs \citep{attwood09}. Thus, the overproduction of BDs in non-radiative simulations substantiates the importance of radiative transfer and feedback from protostars in accurately investigating fragmentation and the initial mass function.

\subsubsection{Accretion Rates}

\begin{figure*}
\epsscale{1.15}
\plottwo{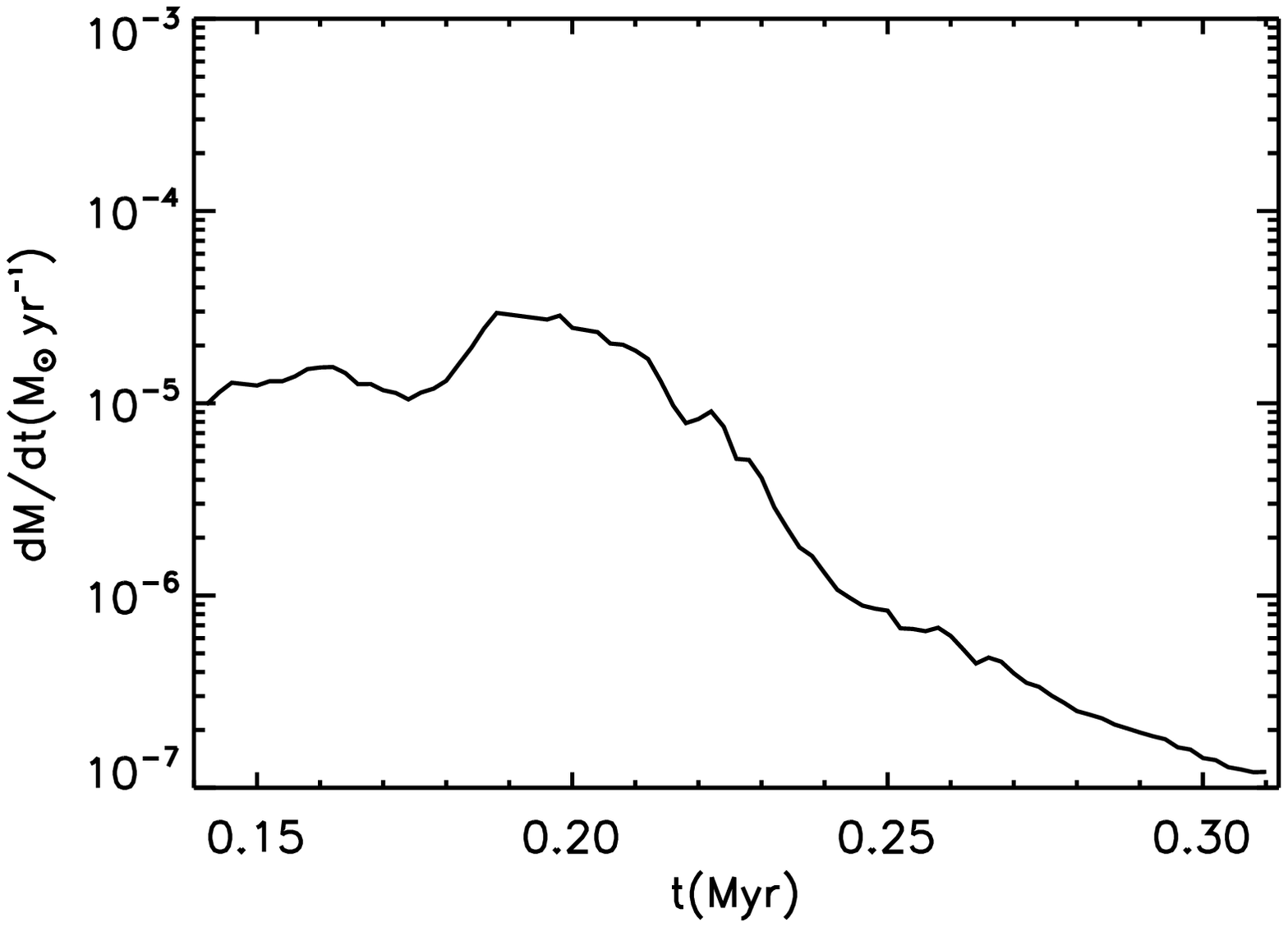}{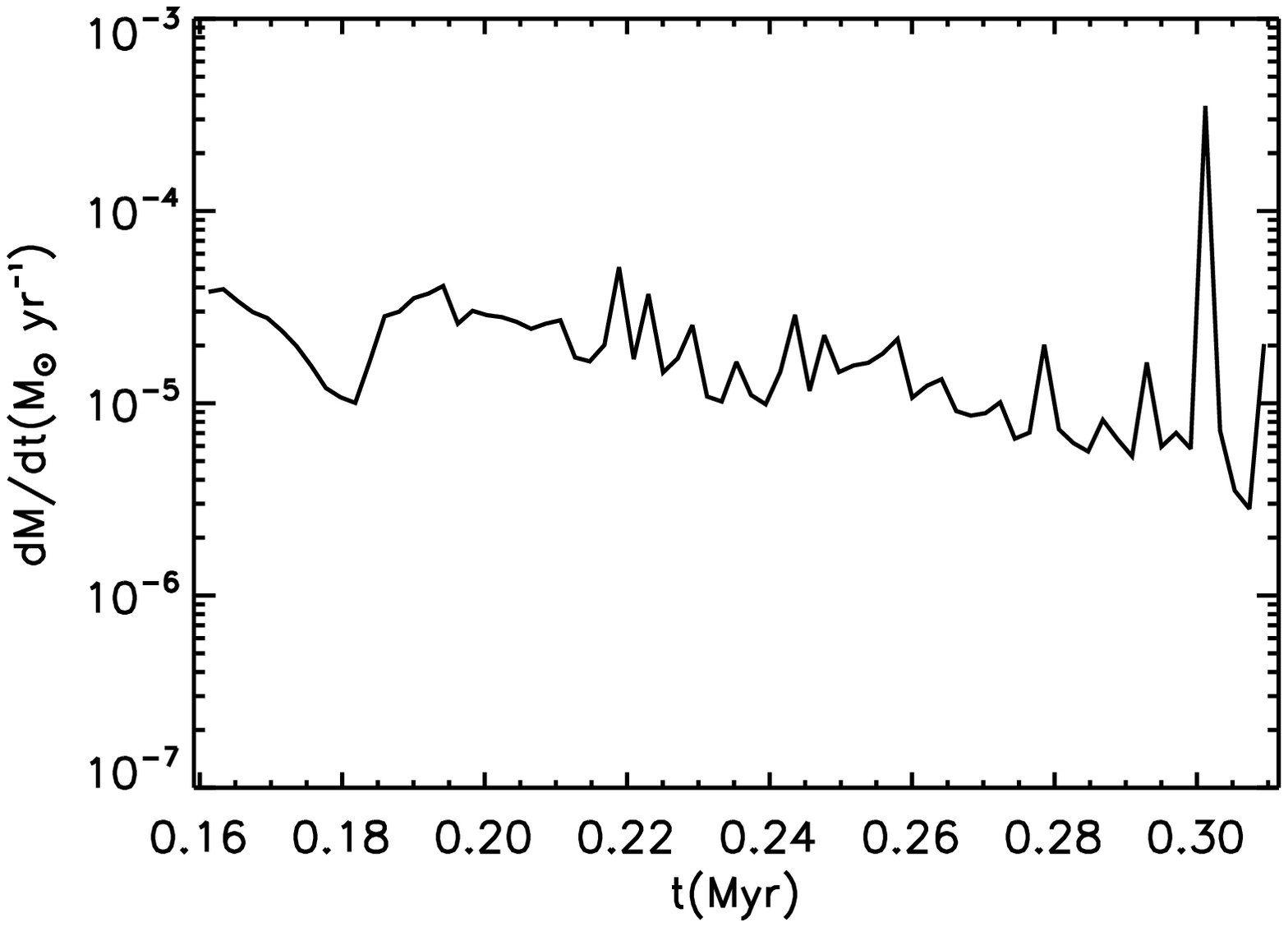}
\caption{The accretion rate, $\dot M$, as a function of time for the first forming object in the RT (left) and NRT (right) simulations. We average both simulations over 2 kyr for consistency.
\label{acc}}
\end{figure*}

As indicated by the Toomre criterion given by equation (5.2), the local gas temperature is key to the stability of disks. Clumpiness in the disks is directly reflected in the variability of the protostellar accretion rate. Figure \ref{acc} shows the accretion rates for the two first-forming protostars in each calculation as a function of time. 
The RT protostellar accretion in the left panel illustrates that once a protostar has accreted most of the mass in the core envelope, its accretion rate diminishes significantly.
Protostars in both simulations show evidence of variable accretion on short timescales. 
However, the accretion bursts in the NRT simulation may vary by an order of magnitude, while
in the RT case variability is generally only a few.  Disk clumpiness may be magnified due to dynamical perturbations by nearby companions. For the cases shown, the RT protostar is single, while the NRT protostar has several companions. 
Similar variability to the NRT protostellar accretion rate is also observed by \citet{schmeja04}. In their turbulent isothermal runs, \citet{schmeja04} show that the magnitude of the initial particle accretion rate is comparable to our calculations at $\dot m \sim \mbox{few} \times 10^{-5}$ $\msun$ yr$^{-1}$ with variability by factors of 5-10. 
However, the reported accretion rates appear to significantly decrease within 0.1 Myr.

In principle, a sizable amount of the protostellar mass may be accreted during the periods of high accretion. We define a burst as an increase of 50\% in the accretion rate over 1000 years, where mergers of another protostar of mass $m > 0.1$ $\msun$ are excluded. Using this metric, the NRT protostars accrete from 0-13 \% of their mass during the bursts with a median of 5\%. The RT protostars accrete 0.0-9 \% of their mass during bursty accretion with a median amount of 1\%. 
Thus, variable accretion is not significant. Our data analysis is limited by the coarse level time step of $\sim 100$ yrs, so that accretion rate variability on shorter timescales will not be resolved in the analysis. For comparison, \citet{vorobyov06}, modeling the formation and accretion history of a protostar in two-dimensions, find that $> 50 $ \% of the protostellar mass is gained in short intense accretion bursts. In their simulations, accretion occurs smoothly until $t  \simeq 0.15$ Myr, where variability on timescales $< 100$ yrs begins, corresponding to accretion of $\sim 0.05$ $\msun$ clumps.  Although their time resolution is finer, sampling at longer time increments, as in our calculation, is unlikely to miss persistent cyclical variability of four to five orders of magnitude in accretion rate. 
We find that when the stellar mass is about half the final mass, large variability in the RT accretion rates is rare, while it is more common in the NRT case.
RT protostars with ages comparable to $t  \simeq 0.15$ Myr experience the most variable accretion occurring over 1-2 orders of magnitude. However, by this time,  the majority of the envelope mass has been accreted and accretion rates are $\bar{\dot m} \sim 10^{-7}$ $\msun$ yr$^{-1}$, so that accreting significant mass is unlikely.  
%

In \citet{vorobyov09}, the authors demonstrate that simulations with a stiffer equation of state and warmer disk exhibit variability over at most two orders of magnitude. This finding is more consistent with our results, and it supports the differences in accretion we find between the NRT and RT calculations.
However, bursty accretion due to disk instability also depends upon the core rotation and the rate at which mass is fed into the disk from the envelope\citep{vorobyov06, boley09}. Thus, we expect that radiative effects alone cannot completely determine accretion behavior.
Since the disks in our low-resolution calculations are not well resolved, it is possible that we may not be able to resolve the disk clumpiness observed by \citet{vorobyov06}. Their innermost cell is placed at 5 AU, which is comparable to the cell size in our high-resolution runs, however, they adopt logarithm spacing to concentrate cells in the inner region of the disks. We note that their method also includes an approximate treatment of magnetic fields that could influence their results and which we neglect in our calculations. 

\begin{figure}
\epsscale{1.2}
\plotone{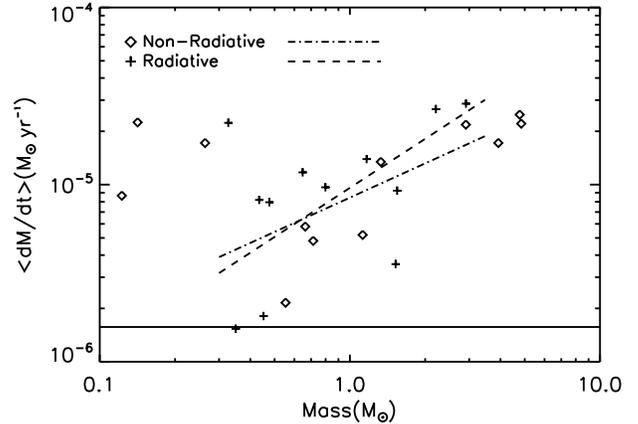}
\caption{The plot shows the distribution of average accretion rates (crosses) as a function of  final star mass at 1 $t_{\rm ff}$. The horizontal line indicates the Shu (1977) accretion rate $c_{\rm s}^3/G$ at 10K.  
The dashed and dot-dashed lines indicate the age weighted fit of the average accretion rates for the RT and NRT runs, respectively.
 \label{aveacc}}
\end{figure}

Figure \ref{aveacc} shows that the NRT simulation exhibits slightly higher average accretion.  
Note that we subtract the accretion spikes caused by significant mergers.
The mean accretion rate over the protostars lifetime for the final protostars is $\sim 1 \times 10^{-5}$ $\msun$ yr$^{-1}$ versus  
$\sim 6 \times 10^{-6}$ $\msun$ yr$^{-1}$ for the RT run.  
Without the added thermal support from radiation feedback and with increased fragmentation, the NRT protostars accrete their envelope mass more quickly.
However, the protostars in both calculations satisfy the same accretion-mass relationship, with accretion increasing approximately linearly with star mass. 
Using a least-squares fitting technique, we obtain power-law relationships $\bar {\dot m} \propto m^{0.92}$ and $\bar {\dot m} \propto m^{0.64}$ for the RT and NRT data, respectively, which have $\chi^2$ values of $67.6$ and $18.0.$\footnotemark We include masses $m \ge 0.1$ \msun in the fit and weight the data by the ages of the protostars. Thus, young protostars with only a short accretion history are weakly weighted. As Figure \ref{aveacc} shows, there is a significant amount of scatter about the fits. 
\citet{schmeja04} find a similar trend between the mean accretion rate and final masses for protostars forming in their isothermal driven turbulence simulations. 
 
 \footnotetext{The $\chi^2$ value for the fit is given by: $\chi^2 = \sum_{i=1}^N {{y_i -A -Bx_i}\over{\sigma_y^2}}$, where $y_i$ are the age-weighted accretion rates, $x_i$ are the masses, $A$ and $B$ are the fit coefficients, and $\sigma_y$ is the standard deviation of the $y_i$ values. }
 
The apparent correlation between stellar mass and average accretion rate occurs because protostars forming in more massive cores tend to be more massive and also have higher accretion rates.
\citet{mckee03} derive a self-similar solution for the accretion rate where the pressure and density each have a power-law dependence on $r$, such that $\rho \propto r^{-k_{\rho}}$ and $P \propto \rho^{\gamma_P}\propto r^{-k_P}$,
where $\gamma_P = 2k_P/(2+k_P)$ and $k_{\rho} = 2/(2-\gamma_P)$. Although the simulated cores are not self-similar, it is possible to fit a power-law to the pressure of the core envelope in most cases. Both RT and NRT cores have exponents in the range $k_P \simeq 0-5$ at a few thousand AU from the protostar, with an average value of $k_P \sim 1$ or $k_{\rho}=1.5$. \citet{mckee03} show that the accretion rate is then:
\begin{eqnarray}
\dot m_* &=& 5.5 \times 10^{-6} \phi_*A^{1/8}, k_P^{1/4} \epsilon_{\rm core}^{1/4} \left(\frac{m_{*f}}{1~\msun}\right)^{3/4} \times \nonumber
\\
& & \left( {{P_{s, \rm core} /k_B}\over{10^6 \mbox{ K cm}^{-3}}} \right) \left({{m_*}\over{m_{*f}}}\right)^{3(2-k_{\rho})/[2(3-k_{\rho})]} \mbox{$\msun$yr$^{-1}$},
\end{eqnarray}
where $m_{*f}$ is the final stellar mass, $P_{s, \rm core}$ is the core surface pressure, $\phi_*$ and $A$ are order unity constants describing the effect of magnetic fields on accretion and the isothermal density profile, respectively.
Since we weight the fit by the protostellar age, this selects for the case where $m_* \simeq m_{*f}$. Assuming that $\Sigma_{\rm cl}$ is roughly constant, $\dot m_* \propto m_{*f}^{3/4}$, that is similar to the slopes produced by the least-squares fit.

\subsubsection{Multiplicity}

The number of stars with stellar companions is an important observable that may directly relate to the initial conditions of star-forming regions. Among the population of field stars, most systems are single with the number of systems containing multiple stars increasing as a function of stellar mass \citep{lada06}. Young pre-main sequence stellar populations are observed to contain more multiple systems than field stars suggesting that the multiplicity fraction evolves over time \citep{duchene07}. 
Unfortunately, the initial stellar multiplicity is challenging to directly measure due to the difficulty of resolving close pairs and limited sample sizes \citep{duchene07}.
The two dominant effects influencing multiplicity are fragmentation and N-body dynamics.
While fragmentation in a collapsing core may result in multiple stars, systems with three or more bodies are dynamically unstable, causing higher-order stellar systems to rapidly lose members. 
Multiple stellar systems can also occur via stellar capture, a mechanism most applicable to high-mass stars forming in very clustered environments \citep{moeckel07}. \citet{goodwin05} suggest that that observed higher-order multiple systems are initially members of open stellar clusters rather than arising from the fragmentation of a single core. In general, the number of such systems is observed to be small, with only 1 in every 50 systems in the field having at least four members \citep{duquennoy91}.

\begin{figure}
\epsscale{1.2}
\plotone{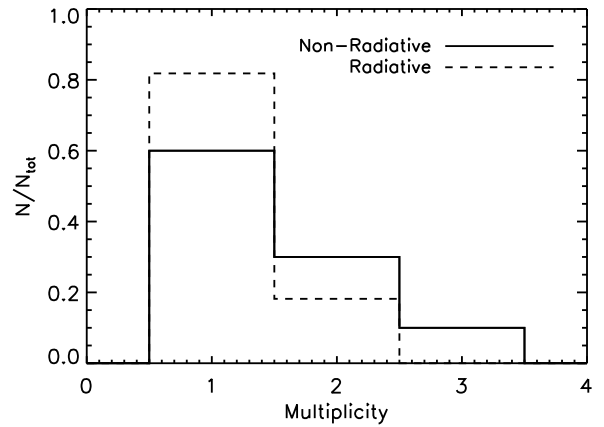} 
\caption{The plot shows the system multiplicity for the two calculations, where N is the number of stellar systems, and the plot is normalized to the total number of systems. The multiplicity on the x-axis is the number of stars in each system.  \label{multiple}}
\end{figure}

The RT and NRT calculations present very different pictures of the initial stellar multiplicity.
The large differences in temperature and fragmentation have a significant effect on the fractions of stars in single and multiple systems. As shown in Figure \ref{multiple}, the majority of stars formed in the RT calculation are single, while in the NRT calculation the majority of stars live in systems with 2 or more stars. This is mainly due to continued disk fragmentation rather than long-lived stable orbital systems. The field single star fraction (SSF), defined as the ratio of the number of primary stars without a {\it stellar} companion to the total number of stellar systems, is observed to be $\sim$ 70\% \citep{lada06}\footnotemark. \footnotetext{The SSF does not include brown dwarfs in estimating multiplicity.} 
The RT calculation produces an SSF of 0.8 $+0.2/-0.4$, 
while the NRT calculation has an SSF of 0.6 $\pm 0.4$, where the uncertainty is given by the poisson error.
Due to the resolution of our calculation, we can only capture wide binary systems of $r > 300$ AU. 
However, a number of protostars have undergone significant mergers, which we define as those in which the smaller mass exceeds 0.1 \msun. 
We find that about a third of the stars in the RT simulation and a tenth of stars in the NRT simulation have
experienced significant past mergers. Assuming that these would have resulted in multiple stellar system revises the SSF values to 0.5$\pm 0.3$ and 0.6$\pm 0.4$, respectively. Unfortunately, this is a very uncertain estimate as we have small statistics, and we cannot know whether the systems with significant mergers would have resulted in bound or unbound systems in the absence of the mergers.

\subsubsection{Stellar Feedback}

Our model includes accretion luminosity and a sub-grid stellar model estimating the contribution from Kelvin-Helmholz contraction and nuclear burning (see Appendix \ref{starpart2}) 
The stellar model includes four evolutionary stages. The earliest stage occurs when the protostar begins burning deuterium within the core at a sufficient rate to maintain a constant core temperature. Once the initial deuterium in the core is depleted, burning occurs at the rate that new matter convects inwards; this is the steady core deuterium state. In the third stage, the star burns the deuterium remaining in the outer layers of the protostar. Finally, the star ceases contracting and reaches the zero-age main sequence (ZAMS).

\begin{figure}
\epsscale{1.2}
\plotone{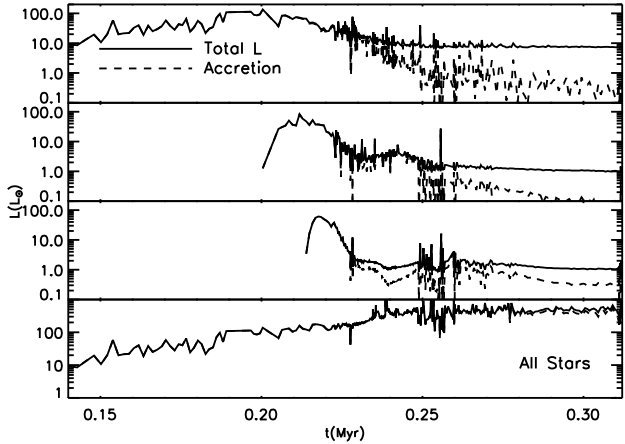} 
\caption{The plot shows the total luminosity as a function of time for three stars in the RT simulation. The accretion luminosity contribution is shown by the dashed line, and the masses are 1.5, 0.45, and 0.35 \msun, respectively.  The bottom plot shows the total luminosity including all the protostars. \label{lumoft1}}
\end{figure}

Figure \ref{lumoft1} shows the luminosity as a function of time for three different protostars. At early times, accretion dominates the luminosity, and variability in accretion is strongly reflected in the total luminosity. At late times, accretion slows and Hayashi contraction begins to make a substantial contribution. 
In general, the total luminosity summed over all the stars is dominated by those protostars with the highest accretion rates. For these young sources, the stellar luminosity is quite small in comparison to the accretion luminosity. Thus, the last panel in Figure \ref{lumoft1} shows that for all times, accretion luminosity is the main source of luminosity.

\begin{figure}
\epsscale{1.2}
\plotone{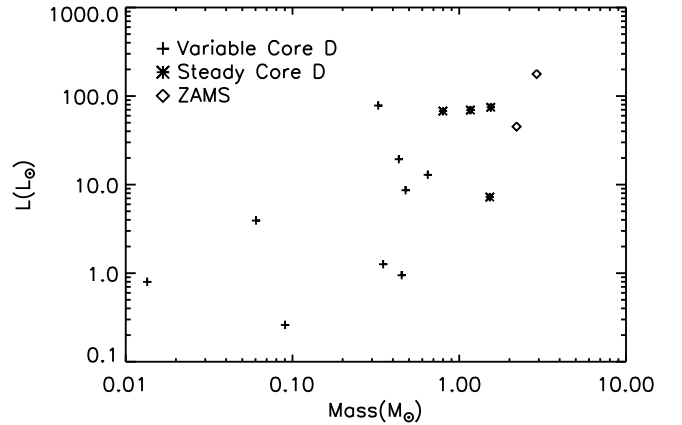}
\caption{ The plot shows the distribution of luminosities (crosses) in the RT simulation as a function of final star mass at 1.0 $t_{\rm ff}$.  The crosses, stars, and diamonds refer to stars undergoing variable core deuterium burning, undergoing steady core deuterium burning, or reaching the zero-age main sequence. \label{avelum}}
\end{figure}

For comparison, luminosity due to other physical processes such as compression and viscous dissipation is small (see Figure \ref{heating}).
Figure \ref{avelum} shows the final luminosity as a function of source mass. The luminosity increases roughly linearly with mass but has a fairly large scatter. 
As indicated on the plot, two of the stars have reached the ZAMS, which was due to increased accretion resulting from significant mergers. Even in this low-mass stellar cluster, there are individual stars with contributions larger than the net viscous dissipation.
This demonstrates that any heating due to viscous dissipation is exceeded by modest protostellar feedback.

\subsection{Resolution and Convergence \label{convergence}}

The AMR methodology allows flexibility in both the depth and breadth of resolution. An insufficient amount of resolution may give inaccurate results, so it is important to gauge the sensitivity of the result to the resolution.  
The large scope of the problem and the expense of the radiative transfer methodology limits the depth or maximum resolution of our calculation, where the RT calculation cost is $\sim$ 70,000 CPU hrs on 2.3 GHz quad-core processors.
 To quantify the effects of resolution on the solution, we run second RT and NRT calculations that evolve the first formed object to a resolution eight times higher than the overall calculation. We run these simulations for 0.12 $t_{\rm ff}$ after the formation of the protostar. We adopt a fixed number of cells for the closest resolved approach between two particles, so that the high-resolution simulations have a merging radius of 32 AU, a factor of eight smaller than the low-resolution cases.

\subsubsection{High-Resolution Study with Radiative Feedback}

\begin{figure*}
\epsscale{1.1}
\plotone{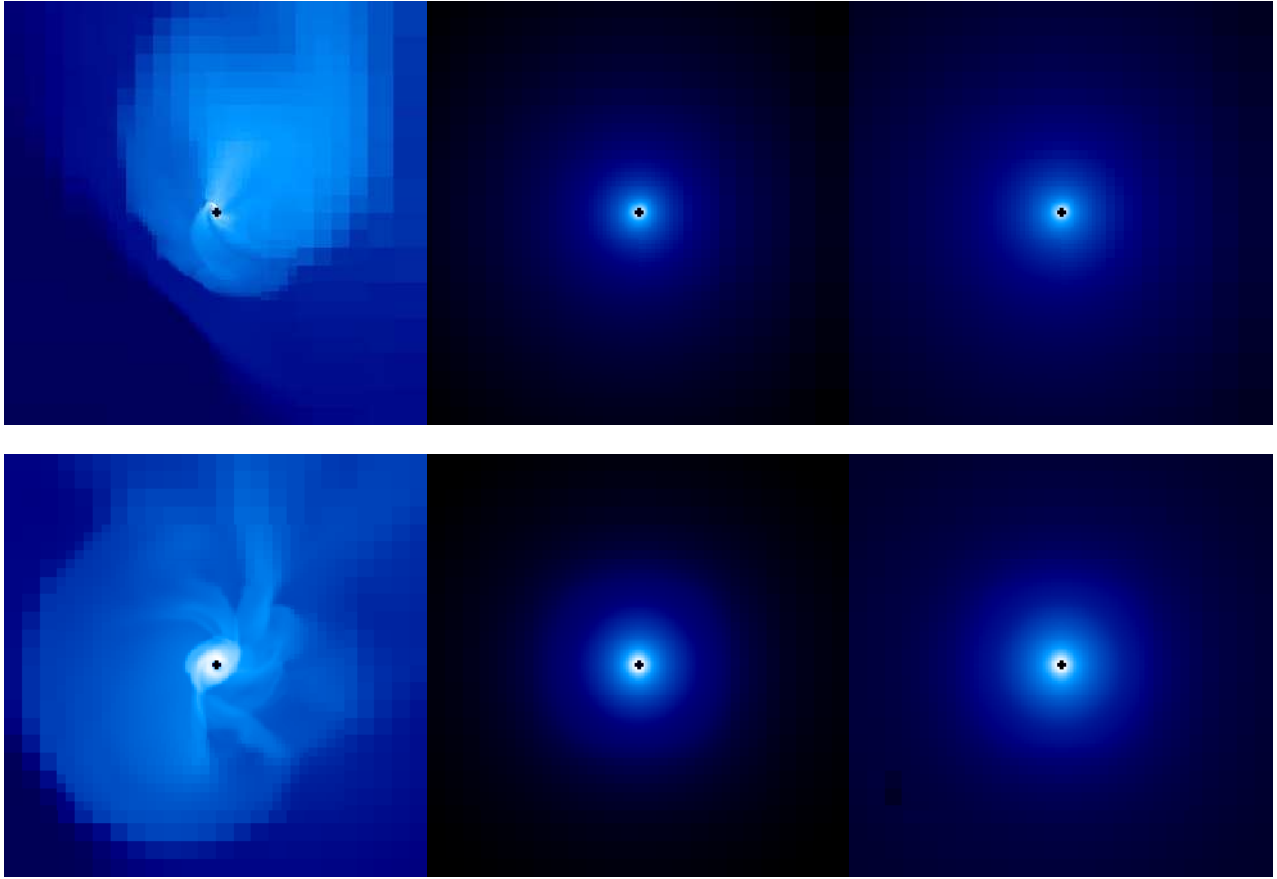} 
\caption{From left to right, the images show the log density, log radiation temperature, $T_{\rm r}=(E_{\rm r}/a)^{1/4}$,  and log gas temperature for a protostellar system at $\sim 0.6~t_{\rm ff}$  followed with $dx = 4$ AU resolution (top) and  $dx = 32$ AU (bottom). The image is 0.03 pc on a side, where we denote the star position with a black cross. The colorscale ranges are given by $10^{-19}-10^{-14}$ g cm$^{-3}$, $1-100$ K, and $1-100$ K, respectively.   \label{tempprof}}
\end{figure*}

The high-resolution and low-resolution calculations both form single objects with stable, thermally supported disks.  Figure \ref{tempprof} shows a comparison of the densities, temperatures, and radiation fields. The effective radiation temperature differs by only a few percent outside the inner cells of the low-resolution calculation. In both cases, the gas and radiation temperatures are well coupled such that $T_{\rm gas} \simeq T_{\rm rad}$. However, the gas in the high resolution case is more centrally concentrated, and the disk radius appears smaller. At the final time, the high-resolution star has accreted 0.54 $\msun$, while the low-resolution case has reached 0.50 $\msun$. 
During the course of the run, the lower resolution case forms a few fragments in the disk, which are almost immediately accreted by the primary, while in the high-resolution case, no additional particles are formed.

\begin{figure}
\epsscale{1.2}
\plotone{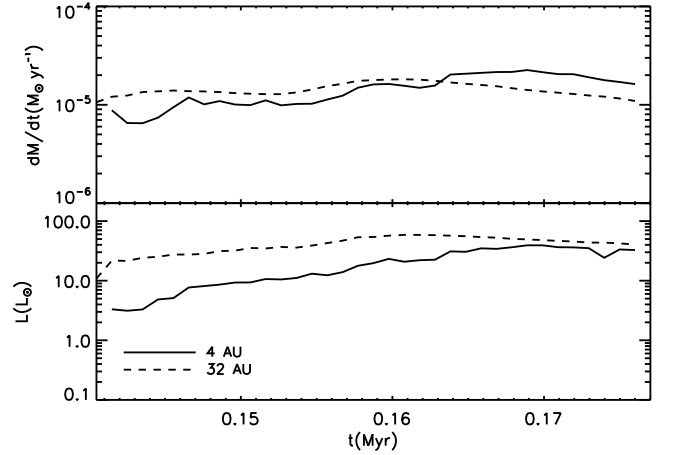}
\caption{The plot shows the accretion rate and luminosity as a function of time for the first formed star in the RT calculation and the same object followed with $dx = 4$ AU resolution. Temporal bins of 1kyr are used. 
\label{lumoft2}}
\end{figure}

Figure \ref{lumoft2} shows a comparison of the accretion and luminosity as a function of time. Accretion is generally smooth, and the rates are generally within a factor of two. The luminosity in the low-resolution run has slightly larger variation, but the two approach a similar value at later times. Although there are deviations in the history between the two runs, the evolution is not significantly different at the higher resolution. Certainly, even higher resolution is preferable for investigation of disk properties, but our main result--{\it that radiative feedback is important to the formation of low-mass stars}--is insensitive to the simulation resolution. 
High-resolution radiation-hydrodynamics simulations of low-mass disks including irradiation confirm that such disks, with properties similar to ours, are stable against fragmentation \citep{cai08}. Gravitational instability is expected to occur only in the regime where the mass of the disks is comparable to the stellar mass \citep{cai08, stamatellos08, stamatellos09}.

\subsubsection{High-Resolution Study with a Barotropic EOS  \label{convergenceEOS}}

\begin{figure}
\epsscale{1.2}
\plotone{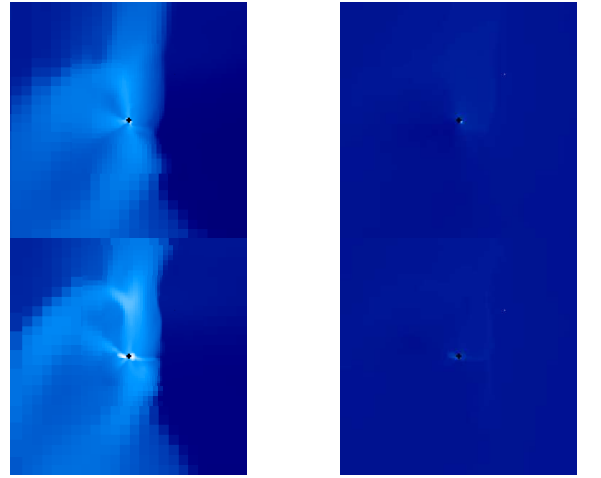}
\caption{  The images show the log density (left)  and log gas temperature (right) for a NRT protostellar system at $\sim 0.5~t_{\rm ff}$  followed with $dx = 4$ AU resolution (top) and  $dx = 32$ AU (bottom). The image is 0.03 pc on a side, where we denote the star position with a black cross. The color scale ranges are given by $10^{-19}-10^{-14}$ g cm$^{-3}$ and $1-50$ K, respectively. \label{barotropic}}
\end{figure}

This higher resolution non-radiative study achieves maximum densities $> 5 \times 10^{-13}$ g cm$^{-3}$, several times higher than the barotropic critical density. Consequently, dense gas is heated to temperatures of $\sim 20-25$ K. 
During the time we compare the non-radiative simulations, both the high-resolution barotropic calculation and the first collapsing core in low-resolution NRT calculation form a similar mass primary object with protostellar disk (see Figure \ref{barotropic}). However, the low-resolution NRT system experiences significantly more fragmentation. We find that the protostellar disk in the NRT case fragments during approximately half of the time steps, while in the higher resolution barotropic case fragmentation occurs very rarely, taking place in less than $<$0.1\% of the time steps.  

Since the low-resolution NRT disks are essentially isothermal, we conclude that heating due to the barotropic approximation is largely responsible for decreasing the number of fragments. In contrast, the higher resolution disks are heated to $\sim 20$ K. However, this is still significantly less heating than in the RT case, and we find that numerical instability is not suppressed completely even with high-resolution. The radiative high-resolution case experiences no disk fragmentation, underscoring our conclusion that radiative feedback is crucial to representing fragmentation or lack thereof in the star formation process.  

Despite different merger radii, in both non-radiative cases all of the fragments are eventually merged with the primary protostar so that the end result in both calculations is a single protostar. This suggests that the fragmentation taking place at low-resolution is largely numerical rather than physical. We emphasize that both significantly higher resolution than we use and additional physics are required to study accretion disk properties.

%

\subsubsection{Convergence}

\begin{figure}
\epsscale{1.2}
\plotone{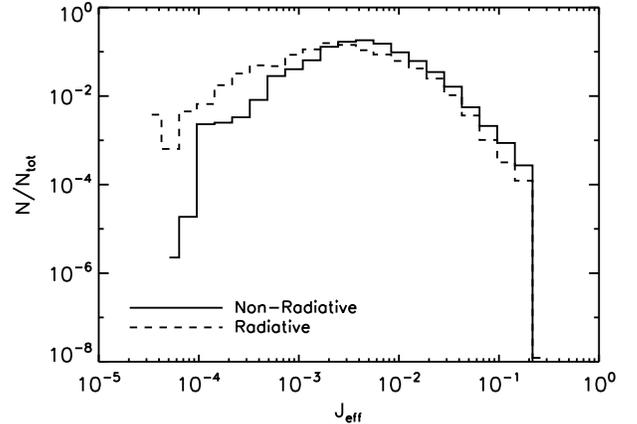}
\caption{Histogram of the effective Jeans number, $J_{\rm eff}$ at  1.0 $t_{ff}$. The solid and dashed lines indicate the NRT and RT simulations, respectively.  Each histogram is normalized to the total number of cells. \label{jeansno}}
\end{figure}

The minimum breadth of resolution is determined by the Truelove criterion.
Due to the radiation gradient refinement criterion we apply to resolve the radiation field, at 1 $t_{\rm ff}$ the RT simulation has $\sim  80$ \% more cells, generally concentrated near the protostars, than the NRT calculation.  This extra refinement improves the resolution regions near protostellar sources. Inverting (5.12) yields an expression for the effective Jeans number for each cell as a function of density, resolution, and sound speed:
\begin{equation}
J_{\rm eff} ={ {(\rho G) ^{1/2} \Delta x_l} \over {c_s \pi ^{1/2} }}.
\end{equation}
As shown in Figure \ref{jeansno}, the RT simulation is shifted to lower $J_{\rm eff}$, where the vast majority of cells in both calculations are resolved to better than $J_{\rm eff}$ = 0.1. 
The choice of base grid resolution guarantees that $J_{\rm eff}$ is typically much smaller than $J$ for most of the cells on the domain. 
We use a fiducial value of $J=0.25$ to trigger additional refinement in both simulations, so no cell has $J_{\rm eff}$ exceeding 0.25. 
Cells in the highest $J_{\rm eff}$ bin are exclusively found on the maximum AMR level, and they are generally at the highest gas densities. These cells, many located in the disks around the protostars, are at the same resolution in both calculations.  Thus, the fragmentation results of the RT and NRT calculations are not dissimilar due to differences in effective resolution but are solely a result of differences in thermal physics.

\section{SIMPLIFYING ASSUMPTIONS  \label{caveat}} 

These numerical calculations neglect a number of arguably crucial physical processes in low-mass star formation. In this section, we discuss the implications for our results.

\subsection{Chemical Processes}

\subsubsection{Dust Morphology}

Our dust model neglects the evolution of dust grains due to coagulation and shattering. In cold dense environments, such as protostellar disks, the aggregation of dust grains may  significantly increase grain 
sizes on timescales as short as 100 years
\citep{schmitt97, blum02}. Observations of Class 0 
protostars indicate significant evolution of the dust size distribution at average densities of $n\simeq 10^7$ cm$^{-3}$ by the Class 0 phase \citep{kwon09}. Since we adopt a single dust model for the entire domain, we are likely to either overestimate or underestimate dust grain size in different regions.
 
To examine the effect of the dust model on gas temperature, we repeat the turbulent driving phase (without gravity) using a conservative model more typical for non-aggregate dust grains:
\begin{eqnarray}
 \kappa_R = 0.015 (T_g^2/110 )  \mbox{ cm$^2$ g$^{-1}$  for $T_g \leq 110$} \label{rosse}\\
 \kappa_P = 0.10 (T_g^2/110 )  \mbox{ cm$^2$ g$^{-1}$  for $T_g \leq 110$}. \label{planck}
\end{eqnarray}
Using this model, we find that shocked gas may be heated as high as 18 K after a crossing time. In comparison, gas in the fiducial case is only heated to $\sim $11 K at the same densities (see Figure \ref{temphist} for the temperature distribution due to the fiducial dust opacity model). However, the extent of the additional heating is quite small since only 0.003\% of the mass is heated above 11  K and thus differs from the fiducial case. This suggests that the simulations may underestimate the gas temperature in low density regions outside of cores ($n_H < 10^{7}$) where the dust distribution is not expected to evolve due to coagulation. Significant discrepancy between the gas temperatures of the two models is mainly confined to a small number of cells and is mitigated by the importance of molecular cooling in these regions, which we discuss in Section 4.1.2.

\subsubsection{Gas Temperature at Low Density} 

To simplify the dust-gas interaction, we assume that dust and gas are perfectly collisionally coupled such that their temperatures are identical.  In molecular clouds, there can be significant variation between the dust and gas temperatures.
For example, dust in close proximity to stellar sources is radiatively heated, while
in strongly shocked regions of the flow, dust acts as a coolant for compressionally heated gas. 
Below we will discuss both the regime where dust cooling dominates, i.e., $T_g > T_d$, and where molecular cooling dominates, i.e., $T_d > T_g$.

When the gas is shock-heated, the perfect coupling approximation remains valid as long as 
the rate of energy transfer between the gas and dust is balanced by the cooling rate of the dust.
The dust cooling per unit grain area by photon emission is:
\begin{equation}
F(a, T_d)= 4 < Q(a,T_d)> \sigma_B T_d^4, 
\end{equation}
where $T_d$ is the dust temperature, $a$ is the grain size, and $<Q(a, T_d)>$ is the Planck-averaged emissivity \citep{draine84}. Then for an ensemble of grains with dust opacity, $\kappa_P$, the dust cooling is given by:
\begin{eqnarray} 
n^2\Lambda_d  &\simeq& 4 \kappa_P \rho \sigma_B T_d^4   \\  
&\simeq& 9 \times 10^{-21} \left({{n_{H}} \over{1.6 \times 10^4 \mbox{ cm}^{-3}}} \right) \left( {T_d \over {10 \mbox{ K}} }\right)^6    \mbox{erg cm}^{-3}\mbox{s}^{-1}. \label{dustcool}
\end{eqnarray}
In equation (\ref{dustcool}), we substitute Equation (\ref{planck}) for $\kappa_{\rm P}$ and assume that $T_g \sim T_d$.
The rate at which energy is transferred from the gas to the dust is given by:
\begin{eqnarray}
n\Gamma_d  &=&  9 \times 10^{-34}  {n_{H}}^2 T_g^{0.5} \left[ 1-0.8e^{ \left( - {{75 \rm K}\over{T_g}} \right)}\right] \times \nonumber \\
& & (T_g-T_d) \left({{\sigma_d} \over{2.44 \times 10^{-21}   \mbox{ cm$^{-3}$}}}\right)  \\ 
&\simeq& 7.3 \times 10^{-24} \left( {  {n_{H}} \over{1.6 \times 10^4 \mbox{ cm}^{-3}} } \right)^2 \left( {T_g \over {10 \mbox{ K}} }\right)^{3/2}  \times \nonumber \\
& &\left[ 1-0.8e^{ \left( - {{75 \rm K}\over{T_g}} \right)}\right]  \left(1 - {T_d \over T_g} \right) \mbox{ erg cm}^{-3} \mbox{ s}^{-1}, \nonumber 
\end{eqnarray}
where we adopt $\sigma_d  = 2.44 \times 10^{-21}$ cm$^{-2}$ for the dust cross section per H nucleus \citep{young04}.
 For a gas temperature of 10 K the exponential term is very small, so we neglect it in the following equation.
Equating these expressions and solving for the gas density at which heating and cooling balance gives:
\begin{equation}
n_{H} \simeq 2 \times 10^7  \left( {T_d \over {10 \mbox{ K}} }\right)^6  \left( {T_g \over {10 \mbox{ K}} }\right)^{-3/2}  \left(1 - {T_d \over T_g} \right)^{-1}   \mbox{cm}^{-3}.
\end{equation}
Thus, we demonstrate that the dust and gas are well coupled as long as the gas densities are sufficiently high.

However, even in regions where the dust and gas may not be well coupled, molecular line cooling is important. 
For gas densities in the range $n_{\rm H} = 10^3-10^5$ cm$^{-3}$, CO is the dominant coolant.
For these densities, the cooling rate per H, $\Lambda/n_{\rm H}$, is approximately constant with density at fixed temperature. To compare to the magnitude of dust cooling consider a 2 km s$^{-1}$ shock that heats the gas above 100 K.  The cooling rate at 100 K is given by
$n^2\Lambda_{\rm mol} \simeq 5 \times 10^{-27} n_{\rm H}$  ergs cm$^{-3}$ s$^{-1}$,
where we adopt the cooling coefficient from \citet{neufeld95}.  The characteristic cooling time is $\sim $ 1000 yrs at the average simulation density, which is approximately half the cell-crossing time of such a shock, implying that molecules cool the gas relatively efficiently. Since the shock temperatures on our domain are limited by our resolution, which is much larger than the characteristic cooling length, post-shock temperatures do not surpass 20 K. In this regime, the dust cooling for perfect dust-gas coupling is at least an order of magnitude larger than the estimated molecular cooling. As a result, we likely under-estimate the temperatures in low-density strongly shocked gas in comparison to similar shocks in molecular clouds. 

In the regions near protostars, the perfect coupling assumption is valid provided that gas heating by dust is balanced by molecular cooling.  This case is discussed by \citet{krumholz08}, where the authors demonstrate that the dust and gas temperature remain within a degree provided that:
\begin{equation}
T_d-T_g \simeq \frac{3.5 \times 10^5}{n_{\rm H}} \mbox{ K}
\end{equation}
for gas temperatures around 10 K. For 
higher gas temperatures around 100 K, we adopt the molecular cooling coefficient above and find that the dust and gas are well coupled provided $n_{\rm H}$ exceeds  $\sim 2 \times 10^8$ cm$^{-3}$.
Number densities of this magnitude are exceeded in collapsing cores, so that regions near protostars are guaranteed to have well-coupled dust and gas. 

Thus, gas temperatures in our RT simulation are fairly accurate for densities larger than the average density, but they may be underestimated by a factor of $\sim 2$ in strong shocks when the molecular cooling rate is much smaller than the implemented dust cooling rate. Since gas heating suppresses fragmentation, our results may actually overestimate the amount of fragmentation. Consequently, our finding that radiative feedback reduces fragmentation would generally be strengthened by a better treatment of the thermodynamics.

\subsection{Magnetic Fields}

Observations indicate the presence of magnetic fields in nearby low-mass star-forming regions \citep{crutcher99}. However, the magnitude of the fields and their importance in the star formation process remain uncertain.  Observations by \citet{crutcher08} suggest that the energy contributed by magnetic fields on core scales is subdominant to the gravitational and turbulent energies. On smaller scales, magnetic fields are believed to be associated with disk accretion and the generation of protostellar outflows \citep{shu94, konigl00}. 

Numerical simulations have demonstrated that the presence of magnetic fields may suppress disk fragmentation by supplying additional pressure support \citep{machida08, price07a, price08}. 
We find that the inclusion of radiative transfer has a similar stabilizing influence on disks.

\subsection{Multi-frequency Radiative Transfer}

Due to the expense of the calculations, we adopt a gray radiative transfer flux-limited diffusion approximation. By averaging over angles and frequencies to obtain the total radiation energy density, we sacrifice the direction and frequency information inherent in the radiation field. As discussed in \citet{krumholz09}, these approximations touch on several competing effects that influence the radiation spectrum. Since radiation pressure is negligible for low-mass stars, it does not affect the gas dynamics. Instead, our main consideration is the extent to which radiative heating may differ for a more sophisticated radiative treatment. As we have discussed in previous sections, the gas temperature and corresponding thermal pressure alone have a significant relationship with accretion and fragmentation.

The first effect to consider is a more exact treatment of dust opacity, which is strongly frequency dependent in the infared and increases towards lower wavelengths (e.g., \citealt{ossen94}).
Since long-wavelength radiation has a lower optical depth, in a multi-frequency calculation the longest wavelengths would be able to escape the core.
Anisotropies in the radiation field may also facilitate cooling.  Radiative beaming, for example via an outflow cavity, may allow photons to escape along the poles \citep{krumholz05}.
Thus, both these effects will likely decrease the temperature in protostellar cores.

The gray radiative transfer method also assumes that the radiation field is thermalized, producing a Planckian radiation spectrum everywhere. 
Although this is a fair assumption in opaque regions where the number of mean-free-paths is large, it fails in optically thin regions.
Since thermalization softens the radiation spectrum, the assumed Planck spectrum is likely to under-predict the heating rate.

Since the net affect of the approximations is somewhat unclear, comparison with more sophisticated radiative treatments would be ideal. However, there have been no 2D or 3D non-gray simulations of low-mass star formation. To date, the most thorough investigation of protostellar formation is presented by \citet{masunaga00}. These spherically symmetric simulations follow the formation of  0.8 and 1.0 $\msun$ protostars. At radii of 60 AU, they find temperatures ranging from 20-250 K during the main accretion phase, while we find $T_{\rm max} \sim 90$ K.  Their maximum protostellar luminosity is 25 $L_{\odot}$, which is entirely due to accretion.  A few of the protostars in the RT simulation have higher masses and higher maximum luminosities, but the gas temperature distributions on average appear similar (see Figure \ref{tempvsr}). However, the disparity in maximum temperature may be attributable to either differences in the radiation schemes or initial conditions and geometry.   
Future work will investigate the effects of 3D multi-frequency radiative transfer on low-mass  star formation.

\section{CONCLUSIONS \label{conclude}}

We perform gravito-radiation-hydrodynamic simulations to explore the effect of radiation feedback on the 
process of low-mass star formation. We compare our calculation with a similar one using an approximately isothermal equation of state in lieu of radiative transfer.  We find that the inclusion of radiative transfer has a profound effect on the gas temperature distribution, accretion, and final stellar masses. 

We confirm the finding of \citet{bate09b} that additional heating provided by radiative transfer stabilizes protostellar disks and suppresses small scale fragmentation that would otherwise result in brown dwarfs. 
However, we also find that the vast majority of the heating comes from protostellar radiation, rather than from compression or viscous dissipation. Thus calculations that neglect radiative feedback from protostars, either because they use approximations for radiative effects that are incapable of including it (e.g., \citealt{bate03, clark08}) or because the explicitly omit it (e.g., \citealt{bate09b}), significantly underestimate the gas temperature and thus the strength of this effect. More generally, we find that, due to significant variations in the temperature with time, no scheme that does not explicitly include time-dependent protostellar heating is able to adequately follow fragmentation on scales smaller than $\sim$0.05 pc.

We find that due to the increased thermal support in the protostellar disks, accretion is smoother and less variable with radiative feedback. However, we show that for low-mass star formation the heating is local and limited to the volume within the protostellar cores. As a result, pre-existing sources do not inhibit turbulent fragmentation elsewhere in the domain.

We find that the mean accretion rate increases with
final stellar mass so that the star formation time is only a weak function of mass.
This is inconsistent with the standard
\citet{shu77} picture, but it is qualitatively consistent with the \citet{mckee03} result
for the turbulent core model, where the star formation time varies
as the final stellar mass to the $1/4$ power. 

The magnitude and variability of protostellar luminosity is of significant observational interest. If accretion contributes a substantial portion of the total luminosity emitted by young protostars, then upper limits for protostellar accretion rates can be inferred directly from the observed luminosity. This may give clues about the formation timescale and the accretion process while the protostars are deeply embedded and cannot be directly imaged. In a future paper we will examine the "luminosity problem'' and compare with embedded Class 0 and Class I protostars.
%

Our larger NRT and RT simulations are performed at a maximum resolution of 32 AU, so it is possible that a few of our cores form stars that otherwise would have  become thermally supported or turbulently disrupted in a higher resolution calculation. Thus, higher resolution calculations would be desirable for further work.
Although we find that the inclusion of radiative transfer has a similar impact as magnetic fields on fragmentation and accretion, simulations examining the interplay of magnetic fields and radiative transfer are important.
To asses the accuracy of our radiative transfer approximations, further simulations with multi-frequency treatment in multi-dimensions with improved dust modeling are also necessary.

\acknowledgments{
The authors acknowledge helpful discussions with Andrew Cunningham and Kaitlin Kratter. 
Support for this work was provided by the US Department of Energy at the Lawrence Livermore National Laboratory under contract  B-542762 (S.~S.~R.~O) and 
DE-AC52-07NA27344 (R.~I.~K.) and the NSF grant PHY05-51164  (C.~F.~M. \& S.~S.~R.~O.);  the NSF grant AST-0807739 and NASA through the Spitzer Space Telescope Theoretical Research Program, provided by a contract issued by the Jet Propulsion Laboratory (M.~R.~K.); the NSF grant AST-0606831 (C.~F.~M \& S.~S.~R.~O). Computational resources were provided by the NSF San Diego Supercomputing Center through LRAC grant UCB267; and the National Energy Research Scientific Computer Center, supported by the Office of Science of the US Department of Energy under contract DE-AC03-76SF00098, though ERCAP grant 80325.}

\begin{appendix}
\section{A. The Star Particle Algorithm}
\label{starpart1}

In this appendix we describe the details of our ``star particle" algorithm we use to represent protostars. Appendix  \ref{starpart1} describes how the star particle algorithm functions within the larger ORION code, while Appendix \ref{starpart2} describes the protostellar evolution code that we use to determine the luminosities of our stars. This division is useful because, from the standpoint of the ORION code, a star particle is characterized by only four quantities: mass, position, momentum, and luminosity. The luminosity is determined by the protostellar evolution model outlined in Appendix \ref{starpart2} that is attached to each star particle, but the only output of this model that is visible to the remainder of the code is luminosity.

In a calculation using star particles, we add a set of additional steps to every update cycle on the finest AMR level, so that the cycle becomes
\begin{enumerate}
\item Hydrodynamic update for gas \label{hydro}
\item Gravity update for gas \label{grav}
\item Radiation update, including stellar luminosity \label{radgas}
\item Star particle update
\begin{enumerate}
\item Sink particle update \label{sinkupdate}
\item Stellar model update \label{starupdate}
\end{enumerate}
\end{enumerate}
Steps (\ref{hydro}) -- (\ref{radgas}) are the ordinary parts of the update that we would perform even if no star particles were present. In steps (\ref{hydro}) and (\ref{grav}) star particles have no direct effect (since they do not interact hydrodynamically, and we handle their gravitational interactions with the gas in an operator split manner in step (\ref{sinkupdate}).

In step (\ref{radgas}), star particles act as sources of luminosity, as indicated in equation (\ref{radenergy}). We implement this numerically as follows: let $L_n$ and $\mathbf{x}_n$ be the luminosity and position of star particle $n$. Our code uses the \citet{krumholzkmb07} radiation-hydrodynamic algorithm, in which we split the radiation quantities into those to be handled explicitly and those to be handled implicitly. We therefore write the evolution equation to be solved during the radiation step as
\begin{equation}
\frac{\partial \mathbf{q}}{\partial t} = \mathbf{f}_{\rm e-rad} + \mathbf{f}_{\rm i-rad},
\end{equation}
where $\mathbf{q}=(\rho, \rho\mathbf{v}, \rho e, E)$ is the state vector describing the gas and radiation in a cell, the explicit update vector $\mathbf{f}_{\rm e-rad}$ is the same as in the standard \citeauthor{krumholzkmb07}\ algorithm (their equation 52)\footnote{Note that our notation here differs slightly from that of \citet{krumholzkmb07}, in that we follow the standard astrophysics convention in which $\kappa$ is the specific opacity, while \citet{krumholzkmb07} follow the radiation-hydrodyanmic convention in which $\kappa$ is the total opacity. As a result, any opacity $\kappa$ that appears in the \citet{krumholzkmb07} equations is replaced by $\kappa\rho$ here.}, and the implicit update is modified to be
\begin{equation}
\mathbf{f}_{\rm i-rad} = \left(
\begin{array}{c}
0 \\
0 \\
-\kappa_{\rm P} \rho (4 \pi B - cE)  \\
\nabla\cdot \left(\frac{c\lambda}{\kappa_{\rm R}\rho}\nabla E\right) + \kappa_{\rm P}(4\pi B - c E) + \sum_n L_n W(\mathbf{x}-\mathbf{x}_n).
\end{array}
\right).
\end{equation}
Here $W(\mathbf{x}-\mathbf{x}_n)$ is a weighting function that depends on the distance between the 
location of the cell center $\mathbf{x}$ and the location of the star $\mathbf{x}_n$. The weighting function has the property that the sum of $W(\mathbf{x}-\mathbf{x}_n)$ over all cells is unity, and that $W(\mathbf{x}-\mathbf{x}_n) = 0$ for $|\mathbf{x}-\mathbf{x}_n|$ larger than some specified value. For the computations we present in this paper we use the same weighting function as we use for accretion (equation (13) of \citealt{krumholz04}). However, we have experimented with other weighting functions, including truncated Gaussians, top-hats, and delta functions, and we find that the choice makes very little difference because radiation injected into a small volume of the computational grid almost immediately relaxes to a configuration determined by diffusion. With this modification to $\mathbf{f}_{\rm i-rad}$, our update procedure is the same as described in \citet{krumholzkmb07}.

Step (\ref{sinkupdate}) is the ordinary sink particle method of \citet{krumholz04}, so we only summarize it here and refer readers to that paper for a detailed description and the results of numerous tests. We first create new particles in any cell whose density exceeds the Jeans density on the maximum AMR level (i.e.,\ where equation (\ref{jeans}) is not satisfied.) Next we merge star particles whose accretion zones, defined to be 4 cells in radius, overlap. This ensures that we combine multiple sink particles created in adjacent cells that simultaneously exceed the Jeans density, or multiple sink particles created in the same cell during consecutive time steps. Then we transfer mass from the computational grid onto existing sink particles. Accretion happens within a radius of 4 cells around each sink particle. The amount of mass that a sink particle accretes is determined by fitting the flow around it to a Bondi flow, reduced to account for an angular momentum barrier that would prevent material from reaching the computational cell in which the sink particle is located. The division of mass accreted among cells inside the 4-cell accretion zone is determined by a weighting function. The accretion process leaves the radial velocity, angular momentum, and specific internal energy of the gas on the computational grid unchanged (in the frame co-moving with the sink particle), and it conserves mass, momentum, and energy to machine precision. Next we calculate the gravitational force between every sink particle and the gas in every cell using a direct $1/r^2$ force computation (since the number of particles is small), and modify the momenta of the sink particles and the momenta and energies of the gas cells appropriately. Finally we update the positions and velocities of each sink particle under their mutual gravitational interaction, using a simple N-body code. Forces are again computed via direct $1/r^2$ sums.

Once the sink particle update is complete, we proceed to update the protostellar evolution model that is attached to each star particle.

\section{B. Protostellar Evolution Model}
\label{starpart2}

Step (\ref{starupdate}) of the update cycle described in Appendix \ref{starpart1} involves advancing the internal state of each star particle. The primary purpose of this procedure is to determine the stellar luminosity for use in step (\ref{radgas}). We determine the luminosity using a simple one-zone protostellar evolution model introduced by \citet{nakano95} and extended by \citet{nakano00} and 
\citet{tan04}. The model has been calibrated to match the detailed numerical calculations of \citet{palla91, palla92}, and it agrees to $\sim 10\%$. The numerical parameters we use for the calculations in this paper are based on this calibration, but we note that after we began this work  \citet{hosokawa09a} published calculations suggesting that slightly different values would improve the model's accuracy. We recommend that \citeauthor{hosokawa09a}'s values be used in future work.

Before diving into the details of the numerical implementation, it is helpful to give an overview of the model. The model essentially treats the star as a polytrope whose contraction is governed by energy conservation. The star evolves through a series of distinct phases, which we refer to as ``pre-collapse", ``no burning", ``core deuterium burning at fixed $T_c$", ``core deuterium burning at variable $T_c$", ``shell deuterium burning",  and ``main sequence". The ``pre-collapse" phase corresponds to the very low mass stage ($m \la 0.01$ $\msun$) when the collapsed mass is not sufficient to dissociate H$_2$ and produce second collapse to stellar densities \citep{masunaga00}. During this phase the object is not yet a star. ``No burning" corresponds to the phase when the object has collapsed to stellar densities, but has not yet reached the core temperature $T_c\approx 1.5\times 10^6$ K required to ignite deuterium, and its radiation is powered purely by gravitational contraction. During this phase the star is imperfectly convective. The next stage, ``core deuterium burning at fixed $T_c$", begins when the star ignites deuterium. While the deuterium supply lasts, core deuterium burning acts as a thermostat that keeps the core temperature fixed and the star fully convective. Once the deuterium is exhausted, the star begins the ``core deuterium burning at variable $T_c$" phase, during which the core temperature continues to rise. The star remains fully convective, and new deuterium arriving on the star is rapidly dragged down to the center and burned. The rising core temperature reduces the star's opacity, and eventually this shuts off convection in the stellar core, beginning the ``shell deuterium burning" phase. At the start of this phase the star changes to a radiative structure and its radius swells; deuterium burning continues in a shell around the radiative core. Finally the star contracts enough for its core temperature to reach $T_c\approx 10^7$ K, at which point it ignites hydrogen and the star stabilizes on the main sequence, the final evolutionary phase in our model.

In the following sections, we give the details of our numerical implementation of this model.

\subsection{Initialization and Update Cycle}

When a star is first created, its mass is always below $0.01$ $\msun$ and thus in the ``pre-collapse" state. We do not initialize our protostellar evolution model until the mass exceeds $0.01$ $\msun$ -- prior to this point star particles are characterized only by a mass and have zero luminosity. On the first time step when the mass exceeds $0.01$ $\msun$, we change the state to ``no burning". Thereafter each star particle is characterized by a radius $r$, a polytropic index $n$, and a mass of gas from which deuterium has not yet been burned, $m_d$. We initialize these quantities to
\begin{eqnarray}
r & = & 2.5 R_{\odot} \left(\frac{\Delta m/\Delta t}{10^{-5}\,M_{\odot}\mbox{ yr}^{-1}}\right)^{0.2} \\
n & = & 5 - 3\left[1.475 + 0.07\log_{10}\left(\frac{\Delta m/\Delta t}{M_{\odot}\mbox{ yr}^{-1}}\right)\right]^{-1} \\
m_d & = & m,
\end{eqnarray}
where $\Delta t$ and $\Delta m$ are the size of the time step when the star passes $0.01$ $\msun$ and the amount of mass accreted during it. If this produces a value of $n$ below 1.5 or greater than 3.0, we set $n=1.5$ or $n=3.0$. These fitting formulae are purely empirical calibrations designed to match \citet{palla91, palla92}. The choice of $n$ intermediate between $1.5$ and $3.0$ corresponds to imperfect convection.

Once a star particle has been initialized and its state set to ``no burning", during each time step we perform the following operations:
\begin{enumerate}
\item Update the radius and the deuterium mass
\item Compute the new luminosity
\item Advance to the next evolutionary phase
\end{enumerate}
We describe each of these operations below.

\subsection{Evolution of the Radius and Deuterium Mass}

Once a star reaches the ``main sequence" evolutionary phase, we simply set its radius equal to the radius of a zero-age main sequence star of the same mass, which we compute using the fitting formula of \citet{tout96} for Solar metallicity. Before this point we treat the star as an accreting polytrope of fixed index. For such an object, in a time step of size $\Delta t$ during which the star gains a mass $\Delta m$,  the radius changes by an amount $\Delta r$ given by a discretized version of equation (5.8) of \citet{nakano00}:
\begin{eqnarray}
\label{radevolution}
\Delta r &=& 2 \frac{\Delta m}{m} \left(1 - \frac{1-f_k}{a_g \beta} + \frac{1}{2}\frac{d\log\beta}{d\log m}\right) r - 2 \frac{\Delta t}{a_g \beta} \left(\frac{r}{G m^2}\right) \left(L_{\rm int} + L_I - L_D\right)r
\end{eqnarray}
Here $a_g = 3/(5-n)$ is the coefficient describing the gravitational binding energy of a polytrope, $\beta$ is the mean ratio of the gas pressure to the total gas plus radiation pressure in the star, $f_k$ is the fraction of the kinetic energy of the infalling material that is radiated away in the inner accretion disk before it reaches the stellar surface, $L_{\rm int}$ is the luminosity leaving the stellar interior, $L_I$ is the rate at which energy must be supplied to dissociate and ionize the incoming gas, and $L_D$ is the rate at which energy is supplied by deuterium burning.

In this equation we adopt $f_k=0.5$, the standard value for an $\alpha$ disk. For $\beta$, the low-mass stars we discuss in this paper have negligible radiation pressure and so $\beta=1$ to very good approximation. In general, however, we determine $\beta$ and $d\log\beta/d\log m$ by pre-computing
a table of $\beta$ values for polytropes as a function of polytropic index $n$ and mass $m$, and then interpolating within that table. For $n=3$ interpolation is unnecessary and we instead obtain $\beta$ by solving the Eddington quartic
\begin{equation}
P_c^3 = \frac{3}{a} \left(\frac{k_B}{\mu m_{\rm H}}\right)^4 \frac{1-\beta}{\beta^4} \rho_c^4,
\end{equation}
where $P_c$ and $\rho_c$ are the central pressure and density of the polytrope (which are also stored in a pre-computed table as a function of $n$), and $\mu=0.613$ is the mean molecular weight for fully ionized gas of Solar composition.

For the luminosity emanating from the stellar interior we adopt
\begin{equation}
L_{\rm int} = \max\left(L_{\rm ms}, L_{\rm H}\right),
\end{equation}
where $L_{\rm ms}$ is the luminosity of a main sequence star of mass $m$, which we compute using the fitting formula of \citet{tout96} for Solar metallicity, and $L_{\rm H} = 4\pi r^2 \sigma T_{\rm H}^4$ is the luminosity of a star on the Hayashi track, with a surface temperature $T_{\rm H}=3000$ K. For the luminosity required to ionize and dissociate the incoming material we use 
\begin{equation}
L_I  = 2.5~L_{\odot} \frac{(\Delta m/\Delta t)}{10^{-5}\,M_{\odot}\mbox{ yr}^{-1}},
\end{equation}
which corresponds to assuming that this process requires 16.0 eV per hydrogen nucleus. The deuterium luminosity depends on the evolutionary stage. In the ``pre-collapse" and ``no burning" phases, $L_D=0$. In the ``core burning at fixed $T_c$" phase, we set the deuterium luminosity to the value required to keep the central temperature at a constant value $T_c=1.5\times 10^6$ K. This is (equation (5.13) of \citealt{nakano00})
\begin{equation}
L_D = L_{\rm int} + L_I + \frac{G m}{r}\frac{\Delta m}{\Delta t} \left\{1-f_k - \frac{a_g \beta}{2}\left[1 + \frac{d\log(\beta/\beta_c)}{d\log m}\right]\right\},
\end{equation}
where $\beta_c = \rho_c k_B T_c / (\mu m_{\rm H} P_c)$ is the ratio of gas pressure to total pressure at the center of the polytrope. In all subsequent phases, deuterium is burned as quickly as it is accreted, so we take
\begin{equation}
L_D = 15~L_{\odot} \frac{(\Delta m/\Delta t)}{10^{-5}\,M_{\odot}\mbox{ yr}^{-1}},
\end{equation}
which corresponds to assuming an energy release of 100 eV per gram of gas, appropriate for deuterium burning in a gas where the deuterium abundance is $\mbox{D}/\mbox{H}=2.5\times 10^{-5}$. Finally, we update the mass of material that still contains deuterium simply based on $L_D$. The change in unburned mass is
\begin{equation}
\label{deutupdate}
\Delta m_d = \Delta m - 10^{-5}M_{\odot} \left(\frac{L_D}{15~L_{\odot}}\right) \left(\frac{\Delta t}{\mbox{yr}}\right).
\end{equation}

\subsection{Computing the Luminosity}

From the standpoint of the rest of the code, the only quantity of any consequence is the luminosity, since this is what enters as a source term in step (\ref{radgas}). The luminosity radiated away from the star consists of three parts:
\begin{equation}
L = L_{\rm int} + L_{\rm acc} + L_{\rm disk},
\end{equation}
where $L_{\rm int}$ is the luminosity leaving the stellar interior as defined above, $L_{\rm acc}$ is the luminosity radiated outward at the accretion shock, and $L_{\rm disk}$ is the luminosity released by material in traversing the inner disk. These in turn are given by
\begin{eqnarray}
L_{\rm acc} & = & f_{\rm acc} f_k \frac{G m \Delta m/\Delta t}{r} \\
L_{\rm disk} & = & (1 - f_k) \frac{G m \Delta m/\Delta t}{r},
\end{eqnarray}
where $f_k=0.5$ as defined above, and $f_{\rm acc}$ is the fraction of the accretion power that is radiated away as light rather than being used to drive a wind. Although we do not explicitly include a protostellar outflow in this calculation, we take $f_{\rm acc} = 0.5$ so that we do not overestimate the accretion luminosity by assuming that the all the accretion power comes out radiatively rather than mechanically.
Thus, we assume a total radiative efficiency of 75\%. Although this value is consistent with x-wind models \citep{ostriker95}, neither x-wind or disk-wind models definitively constrain the total conversion of accretion energy into radiation, and we treat this as a free parameter.

\subsection{Advancing the Evolutionary State}

The final pieces of our protostellar evolution model are the rules for determining when to change the evolutionary state, and for determining what happens at such a change. Our rules are as follows: if the current state is ``no burning", then at the end of each time step we compute the central temperature by numerically solving the equation
\begin{equation}
P_c = \frac{\rho_c k_B T_c}{\mu m_{\rm H}} + \frac{1}{3} a T_c^4,
\end{equation}
where $P_c$ and $\rho_c$ are determined from the current mass, radius, and polytropic index. If $T_c \geq 1.5\times 10^6$ K, we change the evolutionary state to ``core burning at fixed $T_c$" and we change the polytropic index to $n=1.5$.

If the current evolutionary state is ``core burning at fixed $T_c$", then we check to make sure that $m_d \geq 0$ after we update the unburned deuterium mass with equation (\ref{deutupdate}). If not, then the deuterium has been exhausted and we change the state to ``core burning at variable $T_c$".

If the current state is ``core burning at variable $T_c$", we decide whether a radiative zone has formed by comparing the luminosity being generated by deuterium burning, $L_D$, to the luminosity of a zero-age main sequence star of the same mass, $L_{\rm ms}$. We switch the state to ``shell deuterium burning" when $L_D/L_{\rm ms} > f_{\rm rad} = 0.33$. At this point we also change the polytropic index to $n=3$ and increase the radius by a factor of 2.1, representing a swelling of the star due to formation of the radiative barrier.

Finally, if the state is ``shell burning", we compare the radius $r$ at the end of every time step to the radius of a zero-age main sequence star of the same mass. Once the radius reaches the main sequence radius, we switch the state to ``main sequence", our final evolutionary state.
\end{appendix}

\bibliographystyle{apj}
\bibliography{biblio}

\end {document}